\documentclass[journal]{IEEEtran}
\usepackage[T1]{fontenc}

\usepackage{cite}
\usepackage{amsthm}
\usepackage{amsmath,graphicx}
\usepackage{amssymb}
\usepackage{amsfonts}
\usepackage{graphicx,subfigure}
\usepackage{fancyhdr}  %
\usepackage{cases}
\usepackage{extarrows}
\usepackage{algorithm}
\usepackage{algorithmic}
\usepackage{multirow,tabularx}
\usepackage{mathtools}
\usepackage{xcolor}
\usepackage[english]{babel}
\usepackage{caption}
\usepackage{bm}
\usepackage{graphicx}
\usepackage{epstopdf}
\usepackage{cuted}

\theoremstyle{definition}

\newtheorem{theorem}{Theorem}
\newtheorem{lemma}{Lemma}

\newtheorem{proposition}{Proposition}
\newtheorem{definition}{Definition}

\def\proof{\noindent\hspace{2em}{\itshape Proof: }}
\def\endproof{\hspace*{\fill}~$\square$\par\endtrivlist\unskip}

\begin{document}
	
	
	
	\title{Coverage and Rate Analysis for Integrated Sensing and Communication Networks}

	\author{Xu Gan, Chongwen Huang, Zhaohui Yang, Xiaoming Chen, Jiguang He, Zhaoyang Zhang, \\ Chau Yuen,~\IEEEmembership{Fellow,~IEEE}, Yong Liang Guan,~\IEEEmembership{Senior Member,~IEEE}, M\'{e}rouane Debbah,~\IEEEmembership{Fellow,~IEEE}
		\thanks{X. Gan, C. Huang, Z. Yang, X. Chen and Z. Zhang are with the College of Information Science and Electronic Engineering, Zhejiang University, Hangzhou 310027, China (e-mails: \{gan\_xu, chongwenhuang, yang\_zhaohui, chen\_xiaoming, zhzy\}@zju.edu.cn).
			
			J. He is with Technology Innovation Institute, 9639 Masdar City, Abu Dhabi, United Arab Emirates, (email: jiguang.he@tii.ae).
			
			C. Yuen and Y. L. Guan are with the School of Electrical and Electronics Engineering, Nanyang Technological University, Singapore 639798 (e-mail: chau.yuen@ntu.edu.sg, eylguan@ntu.edu.sg).
			
			M. Debbah is with KU 6G Research Center, Khalifa University of Science and Technology, P O Box 127788, Abu Dhabi, UAE (email: merouane.debbah@ku.ac.ae).}
		
	}

	\maketitle
	
	\begin{abstract}
		Integrated sensing and communication (ISAC) is increasingly recognized as a pivotal technology for next-generation cellular networks, offering mutual benefits in both sensing and communication capabilities. This advancement necessitates a re-examination of the fundamental limits within networks where these two functions coexist via shared spectrum and infrastructures. However, traditional stochastic geometry-based performance analyses are confined to either communication or sensing networks separately. This paper bridges this gap by introducing a generalized stochastic geometry framework in ISAC networks. Based on this framework, we define and calculate the coverage and ergodic rate of sensing and communication performance under resource constraints. Then, we shed light on the fundamental limits of ISAC networks by presenting theoretical results for the coverage rate of the unified performance, taking into account the coupling effects of dual functions in coexistence networks. Further, we obtain the analytical formulations for evaluating the ergodic sensing rate constrained by the maximum communication rate, and the ergodic communication rate constrained by the maximum sensing rate. Extensive numerical results validate the accuracy of all theoretical derivations, and also indicate that denser networks significantly enhance ISAC coverage. Specifically, increasing the base station density from $1$ $\text{km}^{-2}$ to $10$ $\text{km}^{-2}$ can boost the ISAC coverage rate from $1.4\%$ to $39.8\%$. Further, results also reveal that with the increase of the constrained sensing rate, the ergodic communication rate improves significantly, but the reverse is not obvious.
		
	\end{abstract}

	\section{Introduction}
	Integrated sensing and communication (ISAC) has been recognized as one of the key usage scenarios for sixth-generation networks, offering efficient wireless communications and sensing functions within the same system and thus enhancing spectrum efficiency \cite{ISAC1,ISAC1_1}. Armed with ultra-reliable and precise dual-function services, ISAC systems can support many emerging applications such as virtual/augmented reality, smart home and factory automation \cite{application}.
	
	Recently, a rich body of research has grown up around the theme of this new technology \cite{ISAC2,ISAC3,ISAC4,ISAC5,ISAC6,chen,wang}. Among them, many efforts have been dedicated to improving the ISAC performance, and most of them focus on waveform designs and signal processing. However, these optimization frameworks lack a mathematical understanding of ISAC systems to explore the fundamental performance bounds and provide analytical insights. Up to now, there are only a few investigations \cite{ISAC_performance,theo_ISAC1,theo_ISAC2,nearfield} that have delved into basic performance analysis on ISAC systems. Specifically, authors in \cite{ISAC_performance} proposed a systematic classification method for both traditional radio sensing and communication, and summarized the major performance metrics and bounds used in sensing, communication and ISAC, respectively. The work in \cite{theo_ISAC1} investigated the trade-off between Cram\'{e}r–Rao lower bound (CRLB) and communication rate on a subspace of the unified ISAC waveforms. Its main contribution was the derivation of the maximum communication rate under the minimum CRLB constraints, and the minimum CRLB under the maximum communication rate constraint. The work in \cite{theo_ISAC2} primarily explored the communication and sensing mutual information and then designed waveforms to strike a trade-off. However, these studies are constrained to specific scenarios and focused on a single cell or few cells, without considering all possible locations of base stations (BSs) and users or inter-cell interference to provide general performance bounds at a network level.
	
	{To tackle this issue, stochastic geometry can provide a powerful tool for deriving general performance analyses and guidelines for multi-cell networks across various possible topologies. Motivated by these considerations, it is of great value to carry out the stochastic geometry-based analyses on the ISAC networks to develop tractable formulations for key performance metrics.}
	
	\subsection{Related Work}
	Leveraging the tractability of stochastic geometry, several notable results have been obtained in separate communication\cite{mmwave,SG1,SG2,SG3,Gil} and sensing \cite{l-localizability,hearability,SG4,SG5} networks. Specifically for the communication networks, authors in \cite{mmwave} proposed a general framework with a distance-dependent line-of-sight probability function to evaluate the coverage and rate performance in millimeter-wave cellular networks. In \cite{SG2}, authors characterized signal-to-interference-plus-noise ratio (SINR) and a unified approach to conduct analytical expressions for symbol error probability (SER), outage probability, and transmission rate analysis. Moreover, the work in \cite{Gil} proved that the coverage probability and average rate can be compactly formulated as a twofold integral for arbitrary per-link power gain with the aid of the Gil-Palaez inversion formula. 
	
	As for the performance analysis for the positioning networks, it is generally necessary to determine the number of BSs participating in the positioning process. In particular, receiving stronger pilot signals from serving BSs is increasingly beneficial for the positioning process. In view of this consideration, authors in \cite{l-localizability} defined and computed the metric, $L$-localizability, to study the probability of the number of BSs whose signals arrive with sufficient quality to successfully participate in the positioning procedure. Besides, the work in \cite{hearability} studied the conditions of lacking an adequate number of detectable positioning signals from localized devices and then derived accurate analytical expressions for the probabilities of meeting this condition in the noncollaborative and collaborative cases, respectively. On the other hand, the metric, CRLB\cite{crlb,crlb2,crlb3} was typically studied to characterize the accuracy of the positioning networks. Specifically, the works in \cite{SG4} and \cite{SG5} first derived the distribution of $L$-localizability probability and then provided heuristic approximations of the CRLB expressions based on the time-of-arrival and received signal strength (RSS), respectively.
	
	{Most recently, the usage of stochastic geometry has been extended to ISAC scenarios \cite{SG_ISAC1,SG_ISAC2,SG_ISAC3,SG_ISAC4}. The work in \cite{SG_ISAC1} defined the sensing signal-to-noise ratio and capacity as well as computed upper and lower bounds of the coverage and capacity expressions for communication and radar sensing, respectively. In \cite{SG_ISAC2}, the detection performance was derived with multiple access for separate communication and radar sensing in a heterogeneous cellular network. Moreover, the work in \cite{SG_ISAC3} modelled radar exploration and communication service tasks using stochastic geometry and defined the summation as the network throughput. It also provided insights into the optimal resource allocation, such as the duty cycle, radar bandwidth, and transmit power, to maximize the defined network throughput. The work in \cite{SG_ISAC4} characterized the performance of radar range detection and communication coverage probability. Furthermore, authors in \cite{SG_ISAC5} derived tractable expressions of area spectral efficiency to capture communication and sensing performance, and formulated the optimization problem to maximize the network performance. The overall results provide an initial worthwhile exploration of stochastic geometry-based analysis for ISAC networks.}
	
	{However, all these studies \cite{SG_ISAC1,SG_ISAC2,SG_ISAC3,SG_ISAC4,SG_ISAC5} are restricted to the scenarios where the ISAC dual functions are respectively applied to the communication users and sensing targets, and obtained separate communication and sensing performance formulations. Thus, the study on the performance analysis of ISAC networks through stochastic geometry is still limited, and the derivations of theoretical formulations for unified ISAC performance are lacking. To cope with it, another scenario well worthy of stochastic geometry analysis is for ISAC operating on the same users \cite{benefit,gan}, which can achieve the mutual benefits of dual functions effectively. As such, the sensing results can facilitate beam management or tracking for the communication process, and in turn, the channel estimation in the communication process can further enhance the sensing performance. However, the theoretical analysis of these ISAC networks through stochastic geometry is still missing and challenging. This paper bridges this gap by deriving the fundamental limits of unified ISAC networks, which solves the difficulties that the performance of ISAC dual functions is coupled with each other, owing to their coexistence with the same BS deployments and experiencing the same channel fading.}

	\subsection{Contributions}
	In this paper, we propose a generalized stochastic geometry framework to model ISAC networks. Based on this framework, we define and calculate the coverage and ergodic rate of joint and conditional ISAC performance, taking into account the coupling effects in coexistence networks. The main contributions of this paper are summarized as follows. 
	\begin{itemize}
		\item \emph{Generalized stochastic geometry framework for ISAC networks:} We propose a generalized stochastic geometry framework for exploring the coverage and ergodic rate of ISAC networks. This framework establishes a unified paradigm to model ISAC networks, capturing the spatial randomness inherent in multi-cell networks.
		
		\item \emph{Coverage and ergodic rate analysis for ISAC networks:} Based on the proposed framework, we obtain theoretical results for the coverage and ergodic rate analysis for sensing and communication performance under resource constraints. Then, we present theoretical results for the coverage rate of unified ISAC performance, taking into account the coupling effects in coexistence networks. These mathematical formulations offer a comprehensive view of how the network deployment influences the ergodic rate of unified ISAC dual functions. Further, we provide the analytical formulations for evaluating the ergodic sensing rate constrained by the maximum SER, and the ergodic communication rate constrained by the maximum CRLB. These results yield an insightful understanding of the capabilities and limitations of one function while ensuring the worst-case performance of the other function in the coexistence networks.

		\item \emph{Network deployment insights into ISAC coverage and ergodic rate performance:} Numerical results validate the accuracy of derived formulations and confirm that denser networks markedly enhance ISAC coverage rate. For instance, increasing the BS density from $1$ $\text{km}^{-2}$ to $10$ $\text{km}^{-2}$ boosts the ISAC coverage rate from $1.4\%$ to $39.8\%$. Further, results also reveal that with the increase of the constrained sensing rate, the communication rate improves significantly, but the reverse is not obvious.
		
	\end{itemize}

	\section{System Model}\label{model}
	In this section, we introduce the generalized stochastic geometry model and characterize performance metrics for evaluating ISAC dual-function performance. Then, we define and calculate the joint and conditional ISAC coverage and ergodic rate.

	\subsection{Network Model}
\begin{figure}[ht]
		\centering{\includegraphics[scale=0.6]{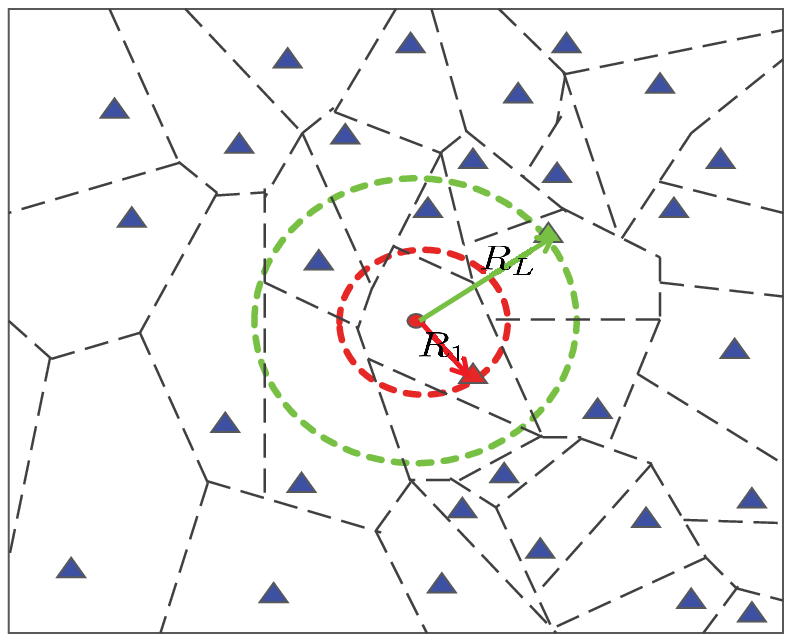}}\vspace{-2pt}
		\caption{An implementation of an ISAC network, where the triangles represent BSs, and the red and green triangles represent the nearest BS and the $\mathcal{L}$-th nearest BS ($\mathcal{L}=4$ in this figure), respectively. The point represents the typical user, and the locations of the remaining users are omitted for brevity.}\label{system}
	\end{figure}

	We consider the ISAC networks where BSs perform the positioning and communication functions on users during downlink transmissions. The locations of BSs are modelled using a homogeneous Poisson point process (PPP)\footnote{{The most widely used model for the spatial locations of nodes is considered as (homogeneous) PPP due to its tractability and analytical flexibility \cite{SG3}. The main advantage of this approach is that the BS positions are all independent which allows substantial analytical tools to be borrowed from stochastic geometry \cite{SG2}.}} $\Phi \in \mathcal{R}^2$ with density $\lambda$ \cite{SG1}. Similarly, $\Phi_u$ represents the homogeneous PPP of users with density $\lambda_u$. The system model is illustrated in Fig.~\ref{system}. {We have summarized all the notations and listed them in Tab.~\ref{tab1} for quick reference.} Generally in the ISAC process, the typical user has multiple access to the positioning pilots from the nearest $\mathcal{L}$ BSs for self-positioning\footnote{{It is assumed that the $\mathcal{L}$ BSs which provide the highest average SINRs make up the set of participating BSs. After averaging the received signal over a period of the positioning process\cite{l-localizability}, the nearest $\mathcal{L}$ BSs are generally selected as in \emph{Definition 2}}}, while receiving the communication data from the nearest BS. Hence, the values of $R_1$ and $R_{\mathcal{L}}$ can characterize the interference boundaries of the ISAC process.
	
	\begin{table}[h!]
		\renewcommand\arraystretch{1.3}
		\begin{center}
			\vspace{3mm}
			\begin{tabular}{ | m{1.3cm} <{\centering}| m{6cm}<{\centering} |}
				\hline 
				\label{table}
				\textbf{Symbols} &\textbf{Definition/explanation}  \\  \hline
				$R_l$ & Distance from the $l$-th nearest BS to the typical user \\ \hline
				$P_T$ & Transmit power of the BS \\ \hline
				$\beta$ & Path-loss exponent (generally $\beta > 2$) \\ \hline
				$\alpha_l$ & Small-scale fading term of the $l$-th link, following exponential distribution with parameter 1 \\ \hline
				$D_l$ &   Directivity gain induced by the directional beam of the $l$-th BS  \\ \hline
				$n_l$ &   Shadowing effect ($n_l \sim \mathcal{N}(0,\xi^2)$)   \\ \hline
				$\gamma$ &  SINR threshold for the positioning process\\ \hline
				$L_P$ &   Maximum number of BSs that can participate in the positioning process    \\ \hline
				$\epsilon_1$ and $\epsilon_2$ & Coverage thresholds for the positioning and communication process, respectively \\ \hline
				$\Gamma(w,z_0,z_1)$ & Generalized incomplete Gamma function as $\int_{z_0}^{z_1} t^{w-1} e^{-t} \mathrm{d} t$ \\ \hline
			\end{tabular}
			\caption{{Notation and symbols used in the paper}}\label{tab1}
		\end{center}
	\end{table}
	
	\begin{lemma}
		Following the above network model, the probability density function (PDF) corresponding to $R_L$ can be written as
		\begin{equation}\label{pdf_rL}
			f_{R_l}(r_l) =   e^{-\lambda \pi r_l^2} \frac{2 (\lambda \pi r_l^2)^{l}}{r_l(l-1)!}.
		\end{equation}
		
		\emph{Lemma 1} can be proved by using the property of PPP distribution. By assigning $L$ the values $1$ and $\mathcal{L}$, the PDFs for $R_1$ and $R_{\mathcal{L}}$ can be obtained, respectively. Moreover, conditioning on a given $R_\mathcal{L}$, the conditional PDF of $R_1$ can be expressed as
		\begin{equation}\label{pdf_r1rL}
			f_{R_1|R_\mathcal{L}}(r_1|R_\mathcal{L}) = \frac{2\mathcal{L}r_1(R_\mathcal{L}^2-r_1^2)^{\mathcal{L}-1}}{R_\mathcal{L}^{2\mathcal{L}}}. 
		\end{equation}
		
		\proof
		One can first calculate the cumulative distribution function (CDF) of $R_L$, which is equivalent to having at least $L$ BSs within $\mathbf{b}(O,r_L)$, where $\mathbf{b}(O,r_L)$ denotes a circle with center at point $O$ and radius $r_L$, i.e.
		\begin{equation}
			F_{R_L}(r_r) = \mathbb{P}(N(r_L) \ge L) = \sum_{j=L}^{\infty} e^{-\lambda \pi r_L^2} \frac{(\lambda \pi r_L^2)^j}{j!},
		\end{equation}
		where $N(r_L)$ is the total number of points within $\mathbf{b}(O,r_L)$. Taking derivation of the resulting CDF with respect to $r_L$ yields Eq.~\eqref{pdf_rL}. Next, it can be proved from \cite{BPP} that the points within $\mathbf{b}(o,R_L)$ obey uniform binomial point process (BPP) under the condition $R_L$, i.e. $\mathbb{P}(N(r_1)=1 | N(R_L)=L) 
		=\begin{pmatrix}
			L \\ 1
		\end{pmatrix} (\frac{r_1^2}{R_L^2})(1-\frac{r_1^2}{R_L^2})^{L-1}$, where $\begin{pmatrix}
			N \\ n 
		\end{pmatrix}= \frac{N!}{n!(N-n)!}$. Taking derivation of the above CDF with respect to $r_1$ yields Eq.~\eqref{pdf_r1rL}.
		\endproof
	\end{lemma}
	
	Using \emph{Lemma 1}, we can obtain the PDFs and conditional PDFs of the locations of the $\mathcal{L}$-th nearest BS and the nearest BS, which are very crucial for deriving the positioning and communication performance subsequently.
	
	\subsection{Performance metrics}
	This section is dedicated to the derivations of performance metrics in ISAC networks. Hence, we derive and calculate the CRLB and transmission rate to measure the ISAC network performance based on the received signals of the typical user.
	
	{Denote the location of the typical user with single antenna and the $l$-th nearest BS with $M$ antennas by $\mathbf{p}_U = [x_U,y_U]$ and $\mathbf{p}_l=[x_l, y_l]$, respectively, and the distance can be written as $r_l = \|\mathbf{p}_U-\mathbf{p}_l\|$. The received signal of the typical user from the $l$-th BS can be written as
	\begin{equation}\label{yl}
		y_l^s = \sqrt{P_T r_l^{-\beta}} \mathbf{h}_l \mathbf{x}_l + z_l
	\end{equation} 
	where $\mathbf{h}_l \in \mathbb{C}^{1 \times M}$ contains small-scale fading coefficients wherein the elements follow $\mathcal{CN}(0,1)$, $\mathbf{x}_l$ is the transmitted vector with $\|\mathbf{x}_l\|^2=1$, and $z_l$ is additive white Gaussian noise. Following the derivation procedure of the CRLB via the RSS method, we first write the logarithm of signal strength from the $l$-th BS based on Eqs.~\eqref{yl} as
	\begin{equation}
		s_{l} = P_{0}^{\text{dB}} - 10 \beta \log(r_l) + n_l   \ (\text{dB}), 
	\end{equation}
	where $P_0^{\text{dB}} = 10 \log(P_T)$ is the transmit power of the BS in dB. Note that the small-scale fading has averaged out over multiple timeslots. The conditional PDF of RSS $f(s_l | \mathbf{p}_U)$ is written as
	\begin{equation}
		\frac{1}{\sqrt{2\pi} \xi} \exp \left(  -\frac{ (s_l - P_{0}^{\text{dB}}+ 10 \beta \log(r_l)   )^2 }{2\xi^2} \right) .
	\end{equation} }
	
	Then the likelihood function of the received vector $\mathbf{s} = [s_1,s_2,...,s_\mathcal{L}]^T $ can be written as $f(\mathbf{s} | \mathbf{p}_U ) = \prod_{l=1}^\mathcal{L} f(s_l | \mathbf{p}_U )$. Thus, the estimated $\mathbf{p}_U$ can be obtained by the maximum likelihood method, i.e., $\mathbf{\hat{p}}_U = \arg \max_{\mathbf{p}_U} f(\mathbf{s} | \mathbf{p}_U) $. The lower bound of the mean square error can be measured by the CRLB, denoted by $\text{CRLB} = \text{Tr}( J_{\mathbf{p}_U}^{-1} )$, where ${J}_{\mathbf{p}_U} = \left( \frac{\partial \mathbf{s}}{\partial \mathbf{p}_U} \right)  J_{\mathbf{s}}  \left( \frac{\partial \mathbf{s}}{\partial \mathbf{p}_U} \right)^T $ and ${J}_{\mathbf{s}} = \mathbb{E} \left\{  \left( \frac{\partial \ln f(\mathbf{s} | \mathbf{p}_U)}{\partial \mathbf{s} } \right) \left( \frac{\partial \ln f(\mathbf{s} | \mathbf{p}_U)}{\partial \mathbf{s} } \right)^T \right\}$. Using algebraic manipulation, we can obtain
	\begin{equation}
		\text{CRLB} = \left( \frac{\ln 10}{10 \beta} \right)^2 \xi^2 \frac{2 \sum_{l=1}^{\mathcal{L}} r_l^{-2} }{\sum_{l=1}^{\mathcal{L}} \sum_{m=1}^{\mathcal{L}} r_l^{-2} r_m^{-2} \sin^2(\theta_l - \theta_m) },
	\end{equation}
	where $\cos\theta_l=\frac{x_U-x_l}{r_l}$ and $\sin\theta_l=\frac{y_U-y_l}{r_l}$.
	
	\begin{proposition}
		Through the adjustment of the orientation or internodal angles of BSs, we can obtain the lower bound of the CRLB as
		\begin{equation}\label{C}
			\underline{\mathcal{C}} = \left( \frac{\ln 10}{10 \beta} \right)^2 \xi^2 \frac{4}{\sum_{l=1}^{\mathcal{L}} r_l^{-2} },
		\end{equation}
		where the lower bound is achieved if it satisfies
		\begin{equation}
			\sum_{l=1}^{\mathcal{L}}  r_l^{-2}  \ge 2 \max\limits_{k \in \{1,...,\mathcal{L}\}}  r_k^{-2}.
		\end{equation}
		
		\proof
		Please see Appendix~\ref{pro1}.
		\endproof
	\end{proposition}
	
	Proposition 1 provides a more tractable form of the positioning performance. It can be noticed in the simulation section that $\underline{\mathcal{C}}$ accurately characterizes the minimum CRLB attainable for the positioning process. Hence, under the condition of $\mathcal{L}$ BSs participating in the positioning process, the positioning coverage is defined as the probability that $\underline{\mathcal{C}}$ is not larger than the threshold value $\epsilon_1$.
	\begin{definition}
		The $\mathcal{L}$-conditional positioning coverage probability is defined as
		\begin{equation}
			P_p(\epsilon_1 \ | \ \mathcal{L}) = \mathbb{P} \left\{ \left( \frac{\ln 10}{10 \beta} \right)^2 \xi^2 \frac{4}{\sum_{l=1}^{\mathcal{L}} r_l^{-2} } \le \epsilon_1  \right\}.
		\end{equation}
	\end{definition}
	
	Then, it is necessary to define the $\mathcal{L}$-localizability probability in order to obtain the marginal positioning coverage rate. In general, the received signal can be successfully detected when its SINR exceeds a predetermined threshold value. Thus, the $\mathcal{L}$-localizability can be defined as the probability that the SINR of the $\mathcal{L}$-nearest BS\footnote{{It is straightforward to infer that $\frac{P_T r_k^{-\beta}}{\sum_{i=\mathcal{L}+1}^\infty P_T r_i^{-\beta} + N_0  } \ge \frac{P_T r_l^{-\beta}}{\sum_{i=\mathcal{L}+1}^\infty P_T r_i^{-\beta} + N_0  }$ for $\forall \ k \le l \le \mathcal{L}$. Therefore, only the SINR of the $\mathcal{L}$-nearest BS is evaluated with respect to $\gamma$.}
	} is greater than the threshold $\gamma$.
	\begin{definition}
		The $\mathcal{L}$-localizability probability is defined to have at least $\mathcal{L}$ BSs participating in the positioning process, written as
		\begin{equation}
			P_L(\mathcal{L} \ | \ \gamma) = \mathbb{P}\left\{ \frac{P_T r_\mathcal{L}^{-\beta}}{\sum_{i=\mathcal{L}+1}^\infty P_T r_i^{-\beta} + N_0  }\ge \gamma \right\}.
		\end{equation}
		where $N_0$ is the average power of received noise.	Note here that the received SINR excludes the small-scale fading and directional beam gain, because it can be assumed that these parts have been averaged out over a period of positioning process and already contained inside $\gamma$\cite{l-localizability}. Then, the probability mass function (PMF) of exact $\mathcal{L}$ BSs involved in the positioning process can be written as
		\begin{equation}
			f_L(\mathcal{L} \ | \ \gamma) = P_L(\mathcal{L} \ | \ \gamma) - P_L(\mathcal{L}+1 \ | \ \gamma).
		\end{equation}
		
	\end{definition}
	
	Combining \emph{Definition 1} and \emph{Definition 2}, the marginal positioning coverage rate can be henceforth deduced as follows.
	\begin{definition}
		The coverage rate in the positioning process is defined as 
		\begin{equation}
			\begin{aligned}
				P_p(\epsilon_1)  = &P_p(\epsilon_1  | L_P) P_L(L_P | \gamma) + \sum_{l=3}^{L_P-1} P_p(\epsilon_1  |  l) f_L(l  |  \gamma)  \\
				&+ u(\epsilon_1-N_L)(1-P_L(3 \ |\  \gamma)).
			\end{aligned}
		\end{equation}
		where $L_P$ is the maximum number of BSs that can participate in the positioning process, the case $l < 3$ corresponds to the unlocalizable scenario, and $N_L$ denotes a sufficiently large value. Specifically, it is impractical for an infinite amount of BSs participating in the positioning process under resource constraints, since the target user needs to have multiple access to the desired pilot signals. Hence, the value of $L_P$ is finite and determined by the resources allocated for the positioning process. Besides, we proceed to consider the scenario where the typical user cannot be positioned if fewer than 3 BSs participated in positioning by the RSS-based method, in which case the corresponding $\underline{\mathcal{C}}$ should take a sufficiently large value $N_L$. In order to ensure the validity of the positioning coverage probability, a step function $u(\cdot)$ is introduced, i.e., $u(\epsilon_1-N_L)=0$ for $\epsilon_1 < N_L$, otherwise 1.
	\end{definition}
	
	{On the other hand, in the communication process, the received signal of the typical user can be expressed as
	\begin{equation}
		y^c = \sum_{l=1}^{\infty} \sqrt{P_T r_l^{-\beta}} \mathbf{h}_l \mathbf{w}_l u_l + n
	\end{equation}
	where $\mathbf{w}_l$ is the beamforming vector of the $l$-th BS, $u_l$ is the transmitted symbol of the $l$-th BS with $|u_l|^2=1$ and $n$ is the additive noise following $\mathcal{CN}(0,\sigma_n^2)$. The typical user is usually associated with the BS that has the smallest path loss. That is, the nearest BS provides the desired signals and the remaining signals are considered as interference. Then, the communication SINR can be written as
	\begin{equation}
		\Upsilon = \frac{P_T r_1^{-\beta} |\mathbf{h}_1 \mathbf{w}_1|^2}{\sum_{l=2}^{\infty} P_T r_l^{-\beta} |\mathbf{h}_l \mathbf{w}_l|^2 + \sigma_n^2}
	\end{equation}
}
	
	{According to the sectored model to approximate directional beam patterns \cite{mmwave}, we have
	\begin{equation}
		D_l = \left\{\begin{matrix}
			M_1,  & \text{with probability}\ c_1 \\
			M_2,  & \text{with probability}\ c_2
		\end{matrix}\right.
	\end{equation}
	where $M_1$ and $M_2$ are main-lobe gain and side-lobe gain, respectively. $c_1 = \frac{\phi}{2\pi}$ and $c_2 = 1-c_1$ are the probabilities of the main lobe and side lobe, respectively, where $\phi$ is the main-lobe width. Then, the communication SINR can be rewritten as
	\begin{equation}
		\Upsilon = \frac{P_T r_1^{-\beta} |\alpha_1|^2 D_1}{\sum_{l=2}^{\infty} P_T r_l^{-\beta} |\alpha_l|^2 D_l + \sigma_n^2},
	\end{equation}
	and the transmission rate can be written as $\mathcal{R} = \log_2(1+\Upsilon)$.}
	
	\begin{definition}
		{The coverage rate in the communication process is defined as
		\begin{equation}
			P_c(\epsilon_2) = \mathbb{P} \{\Upsilon \ge \epsilon_2  \},
		\end{equation}}
	\end{definition}

	Moreover, the coverage rate of ISAC dual functions within the same network needs further definition and calculation.
	\begin{definition}
		The coverage rate of the ISAC process is defined as
		\begin{equation}
			P_{p \& c}(\epsilon_1, \ \epsilon_2) = \mathbb{P} \{ \underline{\mathcal{C}} \le \epsilon_1 , \ \Upsilon \ge \epsilon_2  \},
		\end{equation}
		
		This metric can characterize the probability that the positioning and communication bounds are simultaneously no more than $\epsilon_1$ and $\epsilon_2$, respectively at any point in ISAC networks considering the coupling effects of these two functions.
		
		Meanwhile, the conditional coverage rate of the ISAC process is also desirable, e.g., to find the positioning coverage rate conditioned on communication performance constraints, written as
		
		\begin{equation}
			P_{p  |  c}(\epsilon_1 | \ \epsilon_2)  = \mathbb{P} \{ \underline{\mathcal{C}} \le \epsilon_1 \ |  \  \Upsilon \ge \epsilon_2  \},
		\end{equation}
		which is quite useful, especially when one intends to incorporate positioning capabilities into the underlying communication networks. Specifically, this formulation can directly indicate the achievable positioning coverage rate in the ISAC networks where the accuracy bound of the communication process has been already known. In the other case, we obtain likewise
		\begin{equation}
			P_{c | p}(\epsilon_2 | \ \epsilon_1)  = \mathbb{P} \{  \Upsilon \ge \epsilon_2 \ | \ \underline{\mathcal{C}} \le \epsilon_1  \}.
		\end{equation}
		which describes the coverage of the communication when guaranteeing the positioning performance in ISAC networks. Furthermore, we can accordingly obtain the ergodic rate for positioning, communication, and unified performance.
		
	\end{definition}
	
	\begin{definition}
		The ergodic rates for the communication, positioning, communication constrained by the maximum $\underline{\mathcal{C}}$, and sensing constrained by the maximum $\Upsilon$, can be defined and calculated respectively by
		\begin{equation}
			\mathbb{E}(\underline{\mathcal{C}}) = \int_0^\infty (1 - P_p(\epsilon_1)) \mathrm{d} \epsilon_1, 
		\end{equation}
		\begin{equation}
			\mathbb{E}(\mathcal{R}) = \frac{1}{\ln 2}\int_0^\infty \frac{P_c(\epsilon_2)}{1+\epsilon_2} \mathrm{d} \epsilon_2,
		\end{equation}
		\begin{equation}
			\mathbb{E}(\underline{\mathcal{C}} \ | \ \Upsilon \ge \epsilon_2) = \int_0^\infty (1 - P_{p | c}(\epsilon_1 | \epsilon_2)) \mathrm{d} \epsilon_1,
		\end{equation}
		and
		\begin{equation}
			\mathbb{E}(\mathcal{R} \ | \ \underline{\mathcal{C}} \le \epsilon_1) = \frac{1}{\ln 2} \int_0^\infty \frac{P_{c|p}(\epsilon_2 | \epsilon_1)}{1+\epsilon_2} \mathrm{d} \epsilon_2.
		\end{equation}
		
		\proof
		Please see Appendix-\ref{de6}.
		\endproof	
	\end{definition}
	
	\section{Coverage Probability and Rate Analysis in ISAC Networks}
	This section contains the main technical contributions of this article. In particular, we provide mathematical expressions for the coverage and ergodic rate of positioning, communication and unified performance defined above in the stochastic geometry-based ISAC networks.
	
	\subsection{Coverage Rate Analysis of the Positioning Performance}
	In the following lemma, we derive the coverage rate of positioning performance on the condition of $\mathcal{L}$ BSs participating in the positioning process.
	\begin{lemma}
		The coverage rate  of $\mathcal{L}$-conditional positioning performance is
		\vspace{-5pt}
		\begin{align}\label{le2}\vspace{-5pt}
			&P_p( \epsilon_1   |  \mathcal{L})\approx  1 + \sum_{n=1}^N  \begin{pmatrix}
				N \\ n
			\end{pmatrix} \!\! (-1)^n\!\! \int_{0}^{\infty}\!\!\! \exp\!\big[\! -\! \pi \lambda r_\mathcal{L}^2  \nonumber\\
			&\!+ \!\pi \lambda r_\mathcal{L}^2 e^{-a n \mu r_\mathcal{L}^{-2}} \! \! - \! \pi \lambda \mu \ln\!\left( 1\!-\! e^{-a n \mu r_\mathcal{L}^{-2}} \right)  \big] f_{R_\mathcal{L}}(r_\mathcal{L}) \mathrm{d} r_\mathcal{L},
		\end{align}
		where $a = N(N!)^{-\frac{1}{N}}$, $\mu = \left( \frac{10\beta}{\ln 10} \right)^2 \frac{\epsilon_1}{4\xi^2}$ and $f_{R_\mathcal{L}}(r_\mathcal{L})$ is given in \emph{Lemma 1}.
		
		\proof
		Please see Appendix-\ref{lemma2}.
		\endproof
	\end{lemma}
	
	\emph{Lemma 2} essentially provides analytical expressions of the conditional coverage rate under resource constraints. Further, it is necessary to find the probability distribution of the number of BSs participating in the positioning process, $\mathcal{L}$, which is associated with the received SINR and predefined threshold.
	
	\begin{lemma}
		The $\mathcal{L}$-localizability probability is 
		\vspace{-5pt}
		\begin{align}\label{le3_1}\vspace{-5pt}
			&P_L(\mathcal{L}  | \gamma) = \sum_{n=1}^N  \begin{pmatrix}
				N \\ n
			\end{pmatrix} (-1)^{n+1} \int_0^\infty e^{a n \gamma r_\mathcal{L}^\beta N_0 P_T^{-1}}  \nonumber \\
			& \exp\!\bigg\{\! \pi \lambda r_\mathcal{L}^2\!\left[ \! 1\!-\! e^{-a n \gamma }\!\!+\! (a n \gamma)^{\frac{2}{\beta}} \Gamma(-\frac{2}{\beta}, 0, a n \gamma) \!\right]\!\!\bigg\}\! f_{R_\mathcal{L}}(r_\mathcal{L}) \mathrm{d} r_\mathcal{L},
		\end{align}
		where $\Gamma(w,z_0,z_1)$ is the generalized incomplete Gamma function as $\int_{z_0}^{z_1} t^{w-1} e^{-t} \mathrm{d} t$. Accordingly, the PMF of $\mathcal{L}$ BSs participating in the positioning process is
		\vspace{-5pt}
		\begin{align}\label{le3_2}\vspace{-5pt}
			&f_L(\mathcal{L}  | \gamma)\! =\! \sum_{n=1}^N \begin{pmatrix}
				N \\ n
			\end{pmatrix} (-1)^{n+1}\!\! \int_0^\infty  e^{a n \gamma r_\mathcal{L}^\beta N_0 P_T^{-1}} \exp\big[ \pi \lambda r_\mathcal{L}^2\nonumber \\
			& \big(  1-e^{-a n \gamma }+ (a n \gamma)^{\frac{2}{\beta}} \Gamma(-\frac{2}{\beta}, 0, a n \gamma) \big) \big] - e^{a n \gamma r_{\mathcal{L}+1}^\beta N_0 P_T^{-1}}\nonumber \\
			& \exp\big[ \pi \lambda r_{\mathcal{L}+1}^2 \big(  1-e^{-a n \gamma }+ (a n \gamma)^{\frac{2}{\beta}} \Gamma(-\frac{2}{\beta}, 0, a n \gamma) \big) \big],
		\end{align}
		where $\mathcal{L}$ is restricted to integer values, i.e., $\mathcal{L} \in \mathbb{Z}$.
		
		\proof
		Please see Appendix-\ref{lemma3}.
		\endproof
	\end{lemma}
	
	Now based on \emph{Lemmas 2} and \emph{3}, we are ready to present the theoretical results on the coverage rate of positioning performance.
	\begin{theorem}
		The coverage rate of the positioning performance can be computed as
		\vspace{-5pt}
		\begin{align}\label{p}\vspace{-5pt}
			P_p(\epsilon_1)  =&  P_p(\epsilon_1  | L_P) P_L(L_P | \gamma) + \sum_{l=3}^{L_P-1} P_p(\epsilon_1  |  l)\nonumber\\
			& f_L(l  |  \gamma) + u(\epsilon_1-N_L)(1-P_L(3 \ |\  \gamma)),
		\end{align}
		where $P_p(\epsilon_1  | \mathcal{L})$, $P_L(\mathcal{L} | \gamma)$ and $f_L(\mathcal{L}  |  \gamma)$ are given by Eqs.~\eqref{le2}, \eqref{le3_1} and \eqref{le3_2}, respectively. This corresponds to three scenarios: when $L \ge L_P$, there are total $L_P$ BSs participating in the positioning; when $3\le L < L_P$, the number of involved BSs is determined by the $\mathcal{L}$-localizability SINR threshold $\gamma$; when $L<3$, the typical user is unlocalizable based on RSS methods. 
		
	\end{theorem}
	
	\subsection{Coverage Rate Analysis of the Communication Process}
	We now provide theoretical results of the coverage rate of communication performance considering directional beam patterns.
	\begin{theorem}
		{The coverage rate of the communication performance can be computed as
		\vspace{-5pt}
		\begin{equation}\label{c}\vspace{-5pt}
			\begin{aligned}
				P_c(\epsilon_2) = \int_{0}^\infty \!\! e^{-\frac{\epsilon_2 \sigma_n^2}{P_{T} r_1^{-\beta} M_1 } }  \mathcal{L}_{\mathcal{I}_{agg}}\left( \frac{ \epsilon_2}{P_{T} r_1^{-\beta} M _1 }  \right)  f_{R_1}(r_1) \mathrm{d} r_1 ,
			\end{aligned}
		\end{equation}
		where the Laplace functional of the aggregate interference power is computed as
		\vspace{-5pt}
		\begin{equation}\vspace{-5pt}
			\begin{aligned}
				\mathcal{L}_{\mathcal{I}_{agg}}(s) = & \exp \bigg\{ \pi \lambda r_1^{\beta+2} \left(  1-  \sum_{t=1}^2 \frac{c_t}{1+sP_T M_t}   \right) \\
				&- \pi \lambda \frac{s P_T r_1^2}{2} \bigg[  \sum_{t=1}^2 \frac{c_t M_t}{s P_t M_t + r_1^\beta  } + \frac{2 r_1^{-\beta}}{\beta-2} \\
				& {}_2F_1(1,1-\frac{2}{\beta};2-\frac{2}{\beta};-sP_T M_T r_1^{-\beta}) \bigg]  \bigg\},
			\end{aligned}
		\end{equation}
		and ${}_2F_1(a,b;c;z)$ is the Gauss' hypergeometric function.}
		
		\proof
		Please see Appendix-\ref{theo2}.
		\endproof
	\end{theorem}
	
	\emph{Theorem 2} contains the directional beam-dependent parameters, illustrating the importance of designing the beam gain and width of the main-lobe and side-lobe to improve the coverage rate of communication performance.
	
	\subsection{Joint Analysis in the ISAC Process}
	The coverage analysis of ISAC performance requires taking into account the coupling effects\footnote{{In the proposed ISAC networks, the parameters shared by the dual functions mainly include the BS density, which determines the distances from BSs to the typical user, resource allocation ratio, the transmit power, additive noise power and path loss coefficient.}} in the dual-function coexistence networks, as given in the following theorem.
	\begin{theorem}
		The coverage rate of ISAC performance is
		\begin{align}\label{pc}
			&P_{p \& c} (\epsilon_1, \epsilon_2) = P_L(L_P | \gamma) P_{p \& c}(\epsilon_1, \epsilon_2 | L_p)  \nonumber\\
			& + \sum_{l=3}^{L_P-1}  f_L(l  |  \gamma) P_{p \& c}(\epsilon_1, \epsilon_2 \ | l)  +(1-P_L(3 \ |\  \gamma)) u(\epsilon_1-N_L),
		\end{align}
		where the $\mathcal{L}$-conditional ISAC coverage rate is computed in Eq.~\eqref{theo3_con} and the $\mathcal{L}$-localizability functions $P_L(\mathcal{L} | \gamma)$ and $f_L(\mathcal{L}  |  \gamma)$ are given by Eqs.~\eqref{le3_1} and \eqref{le3_2}, respectively.
		\begin{figure*}[h]
			\begin{small}
				{\begin{align}\label{theo3_con}
					&P_{p \& c}(\epsilon_1,\epsilon_2  |  \mathcal{L})\!\! =\!\! \iint\!\! \bigg\{ \!\! \sum_{n=0}^N \begin{pmatrix}
						N \\ n
					\end{pmatrix} \!\!(-1)^n \!\exp\!\left[\!\frac{\epsilon_2 \sigma_n^2}{P_{T} r_1^{-\beta} M_1 } \! -\! a n \mu r_1^{-2} \!\right]\! \exp \!\! \bigg[\!\!-\! \pi\lambda(r_\mathcal{L}^2 - r_1^2) \sum_{g=1}^G \kappa_g \varrho_g  \bigg( 1\!-\! \sum_{t=1}^2 \frac{c_t}{1+\frac{ \epsilon_2  \varrho_g^{-\beta} M_t}{ r_1^{-\beta} M_1 }} \exp( -a n \mu \varrho_g^{-2}) \bigg)  \bigg] \nonumber\\
					& \exp\bigg\{ \pi \lambda r_\mathcal{L}^2\bigg(  1- \sum_{t=1}^2 \frac{c_t}{1+\frac{ \epsilon_2 M_t}{ r_1^{-\beta} M_1 }} \bigg) - \pi \lambda r_\mathcal{L}^2 \bigg[ \frac{2r_\mathcal{L}^{-\beta }  {}_2F_1\big(1,1-\frac{2}{\beta};2-\frac{2}{\beta}; -\frac{ \epsilon_2 M_t r_\mathcal{L}^{-\beta}}{ r_1^{-\beta} M_1 }\big) }{\beta-2} + \sum_{t=1}^2 \frac{c_t  r_1^{-\beta} M_1 }{\epsilon_2} \bigg] \bigg\} \bigg\} f_{R_1,R_\mathcal{L}}(r_1,r_\mathcal{L})  \mathrm{d} r_1,r_\mathcal{L}.
				\end{align}}
			\end{small}
			\hrulefill
		\end{figure*}
		
		\proof
		Please see Appendix-\ref{theo3}.
		\endproof
		
	\end{theorem}	
	
	Besides, error rate analysis is also an important part included in rate analysis, and the related derivations and simulations can be found in Section.~\ref{error_rate}. 
	
	\section{Numerical Results}\label{simulation}
	In this section, we validate the accuracy of our derived analytical results. Then, we explore the impact of various network parameters on the coverage rate of positioning, communication, and unified performance.
	
	\subsection{Simulation Setup}
	For the simulation setup, it is assumed that the BSs are randomly distributed inside a 2-D plane by a homogeneous PPP with $\lambda=8/\sqrt{3}$ $(\text{km}^{-2})$ \cite{l-localizability}. Parameters for the directional beam patterns include $M_1=0$ dB, $M_2=-20$ dB, and $\phi=30^\circ$. Unless otherwise stated, the simulation parameters are set as follows, $\beta=3.6$, $P_T=0$ dB, $\sigma_n^2=N_0=-89$ dBm, $\xi=-9$ dB and $N=5$. 
	
	\subsection{Coverage and Ergodic Rate Analysis in the Positioning Process}
	\begin{figure}[t]
		\centerline{\includegraphics[width=0.45\textwidth]{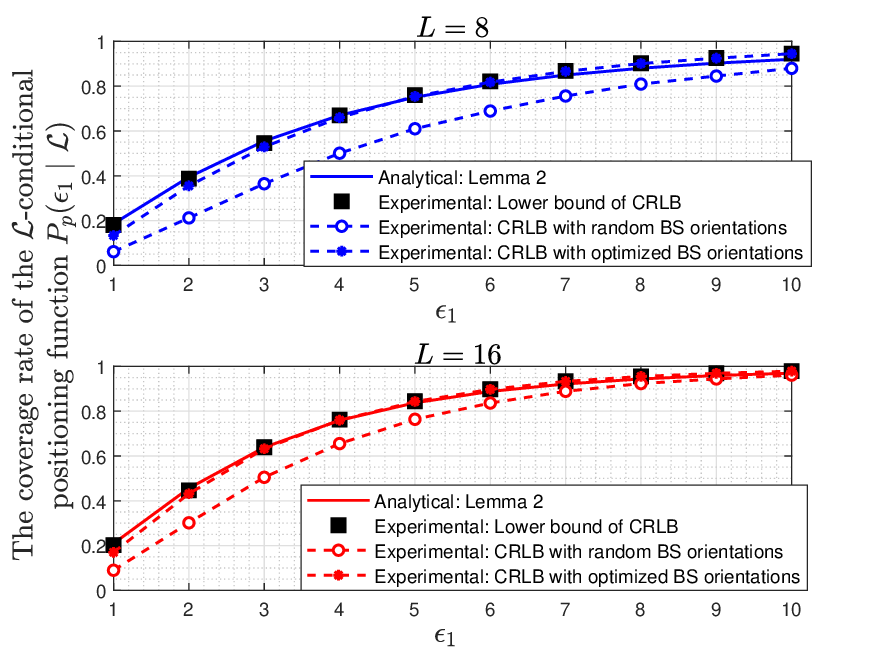}} \vspace{-2mm}
		\caption{The coverage rate of $\mathcal{L}$-conditional positioning function, derived in \emph{Lemma 2}, versus different accuracy thresholds and scenarios.}\label{fig1_theorem1}
	\end{figure}
	
	Fig.~\ref{fig1_theorem1} compares the analytical results of the positioning coverage rate $P_p(\epsilon_1)$ from \emph{Lemma 1} against experimental results across three scenarios. Specifically, the curves in the first scenario are the experimental results for the lower bound of the CRLB derived in \emph{Proposition 1}, which can be found sufficiently close to the analytical results, validating the accuracy of our derived formulations. Then, the performance improvement between the second and third curves suggests that adapting the BS antenna orientations or internodal angles can effectively enhance the positioning performance, especially for high-accuracy positioning. Moreover, it is shown that the CRLB with optimized BS orientations closely matches that of the first curve, which demonstrates the tightness of the derived lower bound.

	\begin{figure}[t]
		\centerline{\includegraphics[width=0.45\textwidth]{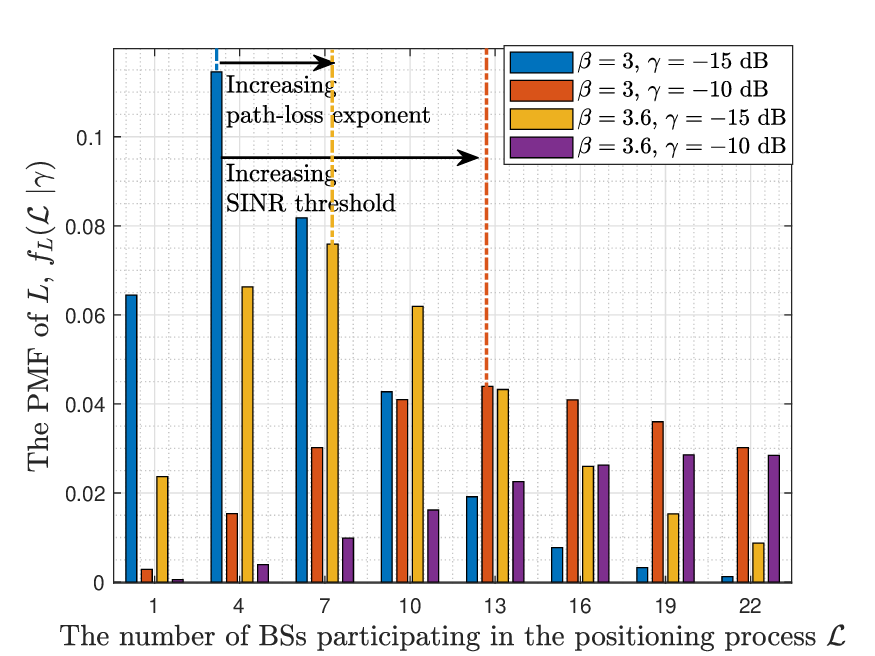}} \vspace{-2mm}
		\caption{The PMF of $\mathcal{L}$, derived in \emph{Lemma 3}, versus different path-loss exponents and $\mathcal{L}$-localizability SINR thresholds.}\label{fig4_theorem2}
	\end{figure}
	In Fig.~\ref{fig4_theorem2}, we describe the PMF of the number of BSs participating in the positioning process, $\mathcal{L}$, and show significant diversity under different $\beta$ and $\gamma$. It is shown that the $\mathcal{L}$ value of the PMF peak becomes larger as $\beta$ and $\gamma$ increase, which can be analyzed by \emph{Lemma 3}. Note that the increase in $\mathcal{L}$ value of the PMF peak does not indicate a higher probability of the larger number of BSs participating in the positioning process. As we can observe, the increases in $\beta$ and $\gamma$ also result in a decrease in the PMF peak values.

	\begin{figure}[t]
		\centerline{\includegraphics[width=0.45\textwidth]{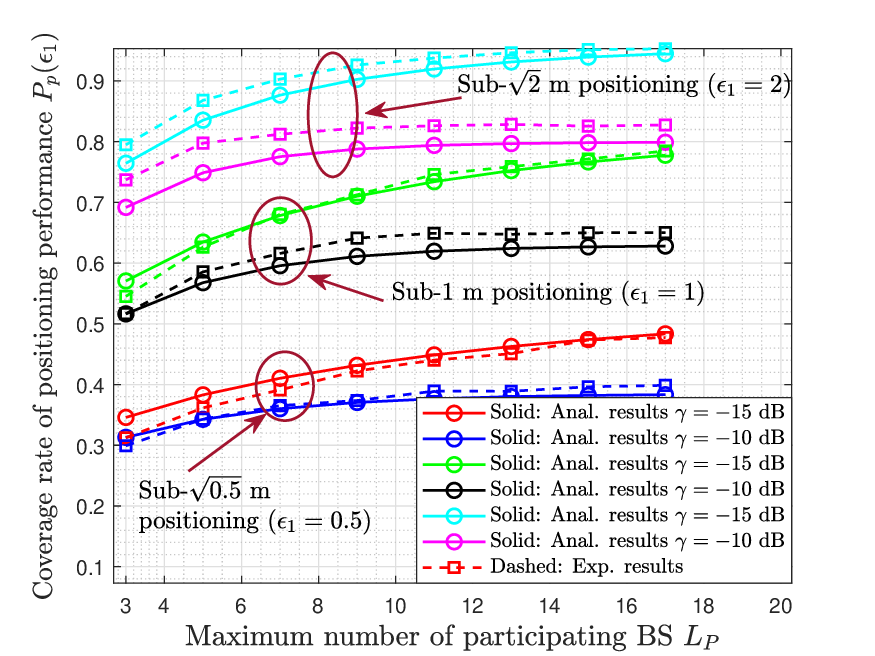}} \vspace{-2mm}
		\caption{The coverage rate of positioning performance $P_p(\epsilon_1)$, derived in \emph{Theorem 1}, versus different accuracy thresholds and $\mathcal{L}$-localizability SINR thresholds.}\label{fig3_theorem3}
	\end{figure}
	In Fig.~\ref{fig3_theorem3}, we explore the relationship between the positioning coverage rate and $\gamma$ under various $L_P$ values. We evaluate the coverage trends for positioning accuracy within $\sqrt{0.5}$ m, $1$ m, and $\sqrt{2}$ m. It shows that increasing $L_P$ significantly improves the positioning performance at small $L_P$ regimes, and reaches saturation at higher values. Therefore, it is not necessary to choose an extremely large $L_P$. In addition, decreasing $\gamma$ can promote the accuracy value even at large $L_P$. For instance, decreasing $\gamma$ from $-10$ dB to $-15$ dB, the coverage rate of sub-$1$ m is improved from $65\%$ to $78\%$.

	\begin{figure}[t]
		\centerline{\includegraphics[width=0.45\textwidth]{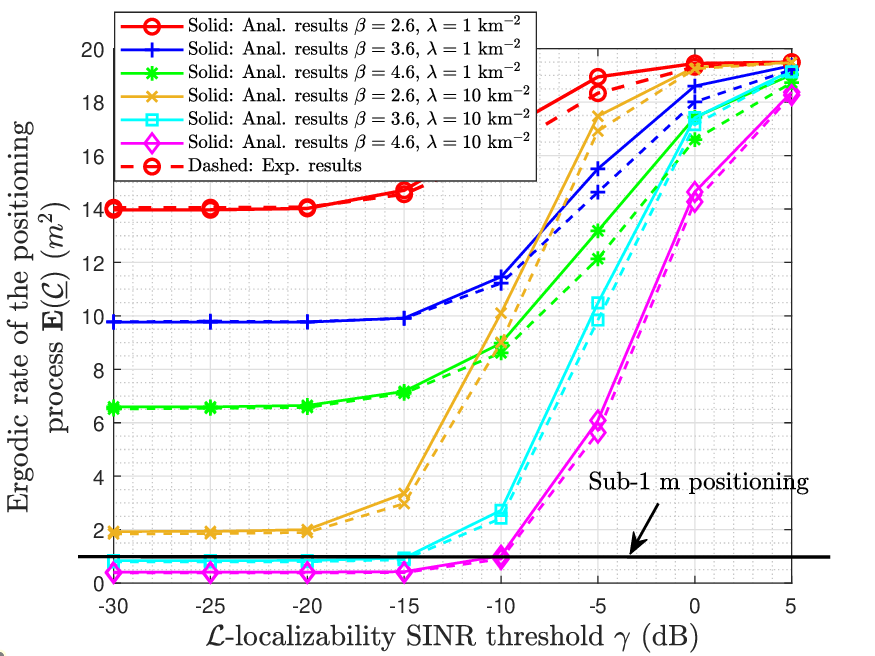}} \vspace{-2mm}
		\caption{The ergodic rate of the positioning process $\mathbb{E}(\underline{\mathcal{C}})$, derived in \emph{Theorem 1} and \emph{Definition 6}, versus different path-loss exponents and BS densities.}\label{fig4_theorem3}
	\end{figure}
	Fig.~\ref{fig4_theorem3} examines the ergodic rate of the positioning process under various $\gamma$ values. This provides an intuitive view of fundamental limits for positioning accuracy. It can be observed that increasing $\beta$ and $\lambda$ properly can improve the positioning performance. Specifically, the ergodic positioning bound is $0.61$ m at $\beta=4.6$, $\lambda=1$ $\text{km}^{-2}$ and $\gamma$ not greater than $-15$ dB. It also presents a clear design guideline for developing desired positioning accuracy, i.e., within $1$ m.
	
	\subsection{Coverage and Rate Analysis in the Communication Process}
	\begin{figure}[t]
		\centerline{\includegraphics[width=0.45\textwidth]{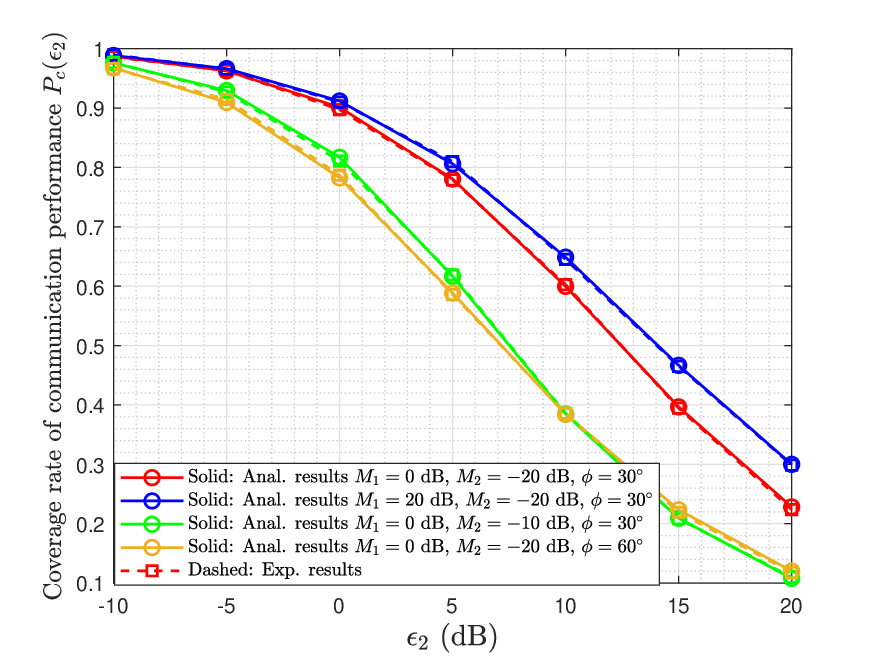}} \vspace{-2mm}
		\caption{{The coverage rate of communication performance $P_c(\epsilon_2)$, derived in \emph{Theorem 2}, under different directional beam settings.}}\label{fig1_theorem4}
	\end{figure}
	{Fig.~\ref{fig1_theorem4} evaluates the impact of directional beam settings on the coverage rate of communication performance. It can be observed that increasing the main-lobe power has some enhancement on the communication coverage, though not significantly, as it simultaneously increases the aggregated interference. On the other hand, widening the main lobe or increasing the power of the side lobe leads to a reduction in communication coverage due to a dramatic increase in interference signal strength. These results suggest that designing a narrow main lobe and the minimum side lobe effects can significantly improve communication coverage.}

	\begin{figure}[t]
		\centerline{\includegraphics[width=0.45\textwidth]{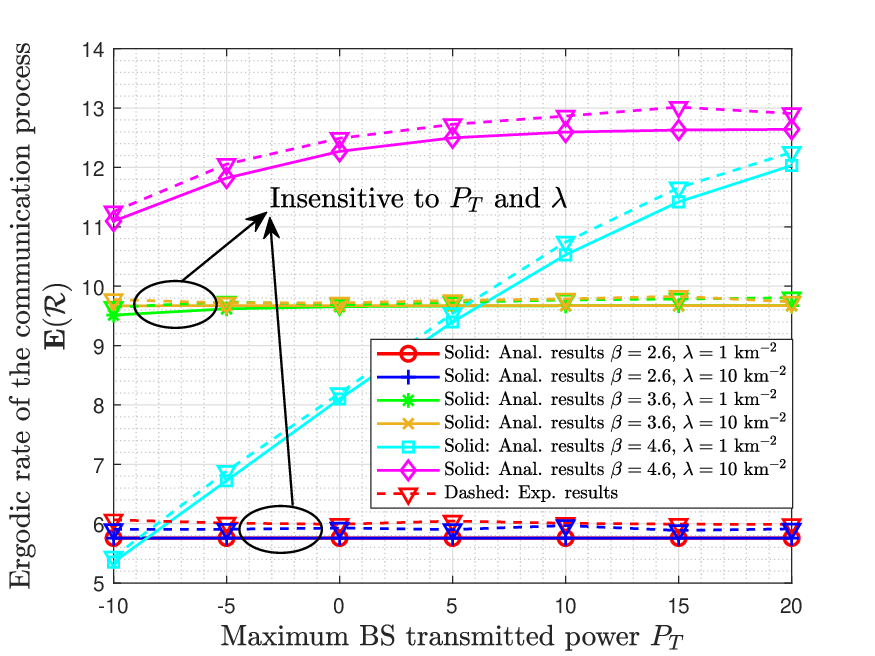}} \vspace{-2mm}
		\caption{{The ergodic rate of the communication process $\mathbb{E}(\mathcal{R})$, derived in \emph{Theorem 2} and \emph{Definition 6}, versus different path-loss exponents and BS densities.}}\label{fig4_theorem4}
	\end{figure}
	{In Fig.~\ref{fig4_theorem4}, we plot the ergodic rate of the communication process under different BS transmitted powers. The results show that the ergodic rate is insensitive to the variations in both $\lambda$ and $P_T$ at lower path-loss exponent regimes. On the other hand, for the scenarios with high path loss, increasing $\lambda$ and $P_T$ can noticeably improve the communication performance, especially for small $\lambda$. This indicates that increasing $\lambda$ and $P_T$ can significantly improve communication coverage, but such changes are marginal under high SINR conditions. These results also provide an intuitive view of the fundamental limits of communication capacity.}
	
	\subsection{Joint Coverage and Rate Analysis in the ISAC Process}
	\begin{figure}[t]
		\centerline{\includegraphics[width=0.45\textwidth]{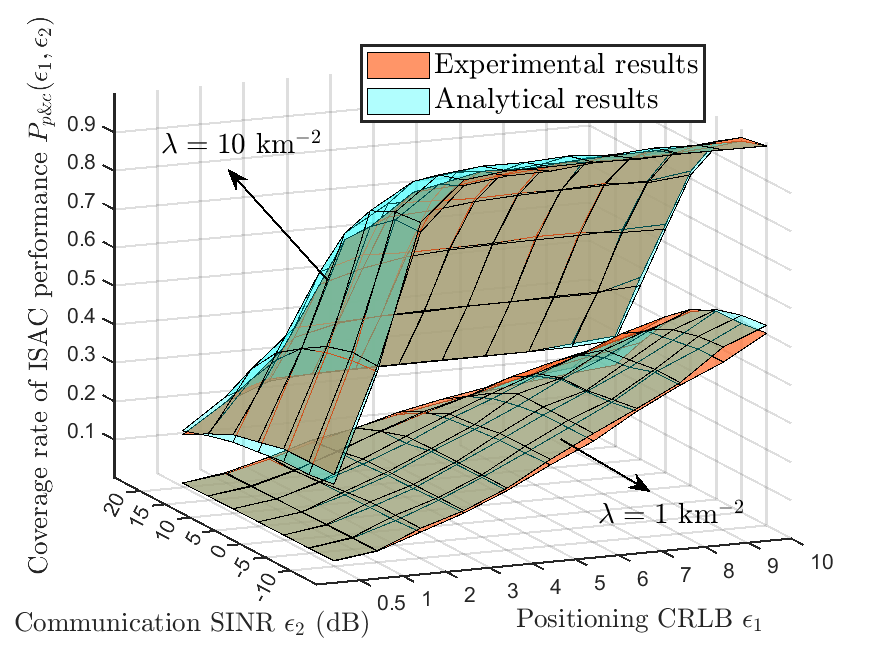}} \vspace{-2mm}
		\caption{{The coverage rate of ISAC performance $P_{p \& c}(\epsilon_1,  \epsilon_2)$, derived in \emph{Theorem 3}, versus different BS densities.}}\label{fig1_3D}
	\end{figure}
	{Fig.~\ref{fig1_3D} depicts the coverage rate of ISAC performance for $\lambda=1$ $\text{km}^{-2}$ and $10$ $\text{km}^{-2}$ cases. The simulation results are observed to overlap the analytical results, validating the accuracy of \emph{Theorem 3}. It also demonstrates that the ISAC coverage improves with the increase of the BS density.}

	\begin{figure}[t]
		\centerline{\includegraphics[width=0.45\textwidth]{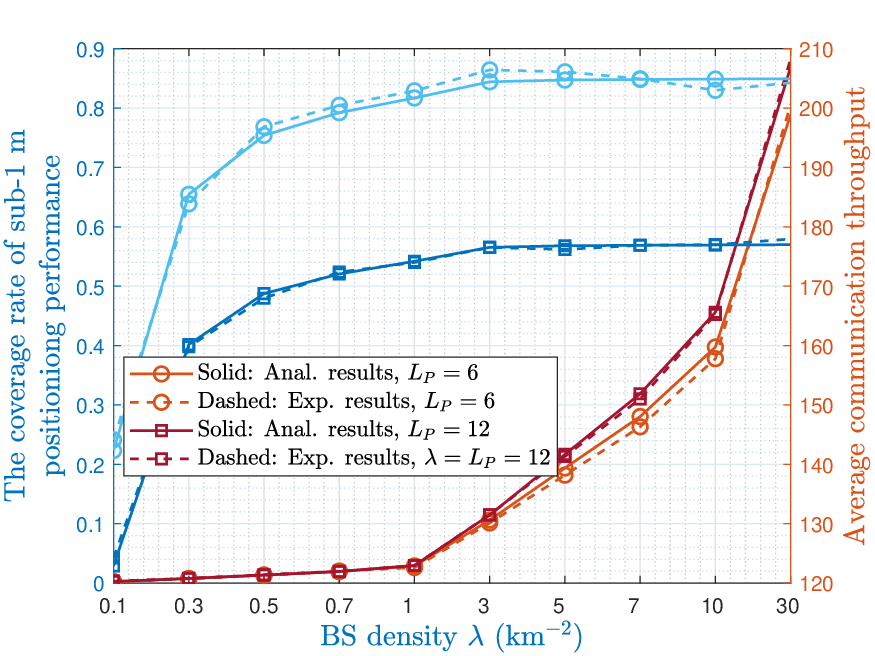}} \vspace{-2mm}
		\caption{{Trade-off between the coverage rate of sub-$1$ m positioning function and average communication throughput, versus different BS densities.}}\label{fig2_other}
	\end{figure}
	{Fig.~\ref{fig2_other} illustrates the trade-off performance between the coverage rate of the sub-1 m positioning function and the average communication throughput varying the BS density with different $L_P$. It can be found that as $\lambda$ increases, both the positioning coverage rate and the communication throughput improve accordingly. The main difference is that the increase in $\lambda$ notably enhances the average communication throughput while having a limited effect on the positioning coverage rate at small $\lambda$ regimes, and that the increase in $\lambda$ can significantly improve the positioning coverage rate while having little effect on the average communication throughput at large $\lambda$ regimes, which demonstrates the optimal $\lambda$ for the communication and positioning performance are not equal. Besides, increasing $L_P$ improves the coverage rate of positioning performance, but severely degrades average communication throughput. These trade-offs highlight the significance of selecting an optimal $\lambda$ and $L_P$ to achieve both high-accuracy positioning and high-throughput communication. }
	
	\subsection{Joint and Conditional Coverage and Error Ergodic Rate Analysis in the ISAC Process}\label{error_rate}
	\begin{figure}[t]
		\centerline{\includegraphics[width=0.45\textwidth]{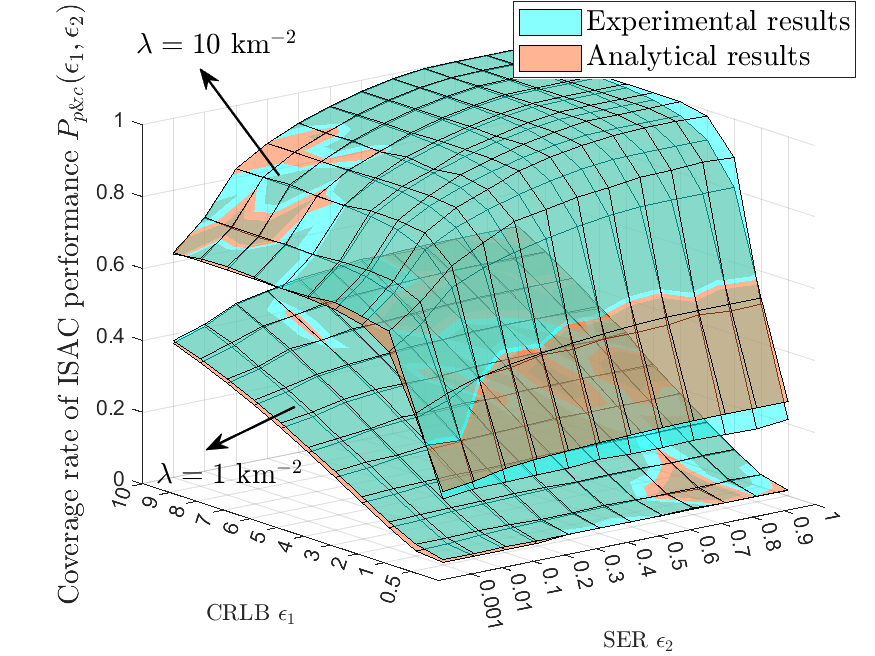}} 
		\caption{The coverage rate of ISAC performance $P_{p \& c}(\epsilon_1,  \epsilon_2)$, derived in Eq.~\eqref{pc2}, versus different BS densities.}\label{fig1_3D_2}
	\end{figure}
	
	\begin{figure}[t]
		\centerline{\includegraphics[width=0.45\textwidth]{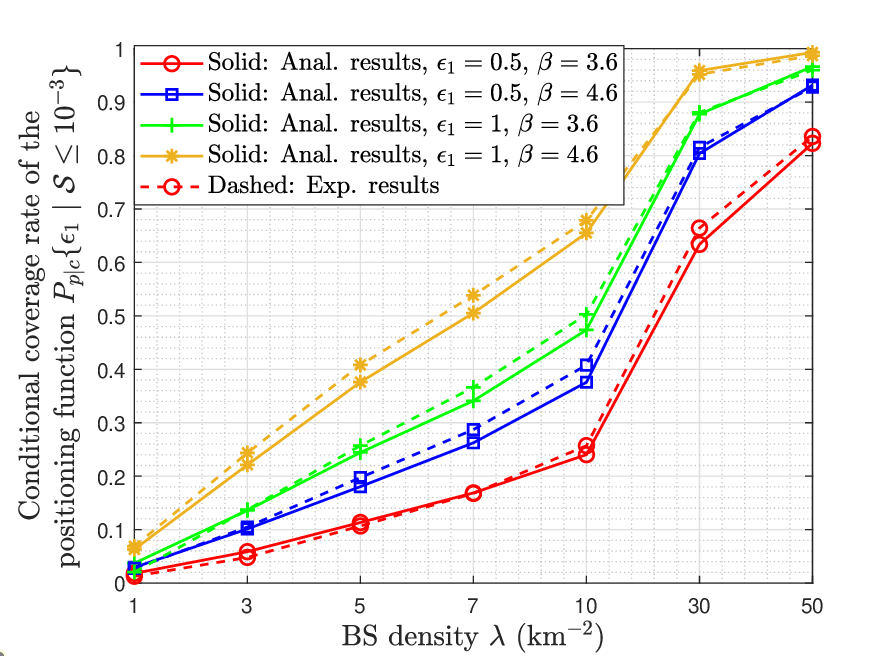}} \vspace{-2mm}
		\caption{The conditional coverage rate of the positioning function constrained by the communication rate, versus different thresholds and path-loss exponents.}\label{fig2_theorem5}
	\end{figure}
	
	\begin{figure}[t]
		\centerline{\includegraphics[width=0.45\textwidth]{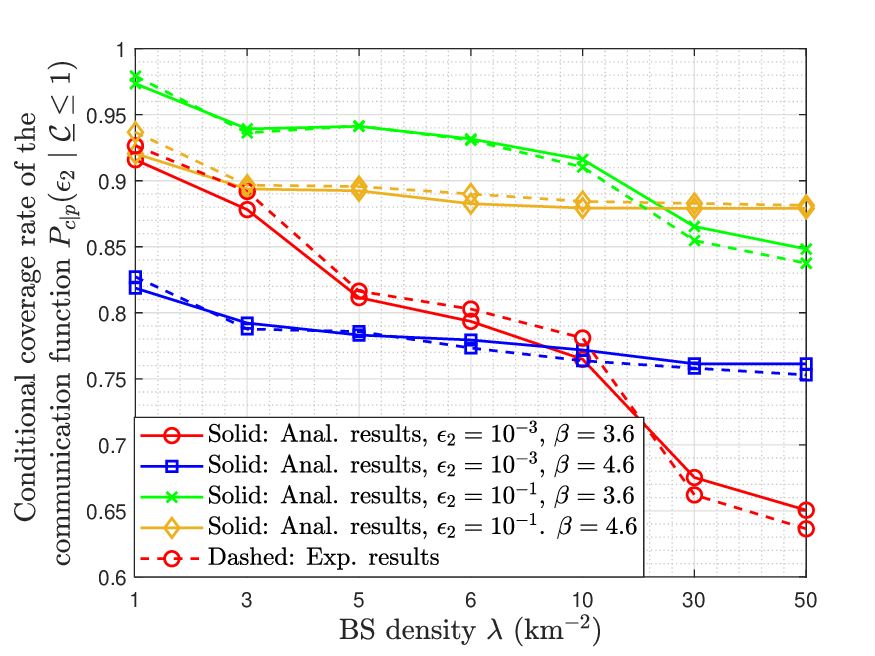}} \vspace{-2mm}
		\caption{The conditional coverage rate of the communication function constrained by the positioning rate, versus different thresholds and path-loss exponents.}\label{fig3_theorem5}
	\end{figure}
	
	In this subsection, we study the SER performance for the communication process, which is derived with a coherent maximum likelihood detector under $K$-quadrature amplitude modulation (QAM) modulation as
	\small
	\begin{equation}\label{S}
		\mathcal{S}\! =\! 4 \frac{\sqrt{K}-1}{\sqrt{K}}  Q\!\left(\!\sqrt{\frac{3}{K-1} \Upsilon}\right)\! -\! 4 \left(\frac{\sqrt{K}-1}{\sqrt{K}} \right)^2\! Q^2\!\left(\!\sqrt{\frac{3}{K-1} \Upsilon}\right).
	\end{equation}
	\normalsize
	
	Note that Eq.~\eqref{S} captures the error probability of a modulated signal going through additive white Gaussian noise channels while not considering the actual coding methods. Then, the coverage rate in the communication process and ISAC process can be defined respectively as
	\begin{equation}
		P_s(\epsilon_3) = \mathbb{P} \{\mathcal{S} \le \epsilon_3  \},
	\end{equation}
	and
	\begin{equation}
		P_{p  \&  s}(\epsilon_1 , \ \epsilon_3)  = \mathbb{P} \{ \underline{\mathcal{C}} \le \epsilon_1 \ ,  \ \mathcal{S} \le \epsilon_3  \},
	\end{equation}
	
	Using similar approaches, we can obtain
	\begin{equation}\label{c2}\vspace{-5pt}
		\begin{aligned}
			P_s(\epsilon_3) =& \int_{0}^\infty e^{-\frac{\sigma_n^2}{P_{T} r_1^{-\beta} M_1 \varsigma} Q^{-2}(\frac{1-\sqrt{1-\epsilon_3}}{2v}) } \\
			&   \mathcal{L}_{\mathcal{I}_{agg}}\left( \frac{ Q^{-2}(\frac{1-\sqrt{1-\epsilon_3}}{2v})}{P_{T} r_1^{-\beta} M _1 \varsigma}  \right) f_{R_1}(r_1) \mathrm{d} r_1 ,
		\end{aligned}
	\end{equation}
	and
	\begin{align}\label{pc2}
		&P_{p \& s} (\epsilon_1, \epsilon_3) = P_L(L_P | \gamma) P_{p \& s}(\epsilon_1, \epsilon_3 | L_p)  \nonumber\\
		& + \sum_{l=3}^{L_P-1}  f_L(l  |  \gamma) P_{p \& s}(\epsilon_1, \epsilon_3 \ | l)  +(1-P_L(3 \ |\  \gamma)) u(\epsilon_1-N_L),
	\end{align}
	where $v=\frac{\sqrt{K}-1}{\sqrt{K}}$, $\varsigma = \frac{3}{K-1}$, $Q^{-2}(\cdot)$ is the square of the inverse Q function, the $\mathcal{L}$-conditional ISAC coverage rate is computed in Eq.~\eqref{theo3_con2} and the $\mathcal{L}$-localizability functions $P_L(\mathcal{L} | \gamma)$ and $f_L(\mathcal{L}  |  \gamma)$ are given by Eqs.~\eqref{le3_1} and \eqref{le3_2}, respectively.
	\begin{figure*}[t]
		\begin{small}
			\begin{align}\label{theo3_con2}
				&P_{p \& s}(\epsilon_1,\epsilon_3 \ | \ \mathcal{L})\! =\! \iint \bigg\{ \sum_{n=0}^N \begin{pmatrix}
					N \\ n
				\end{pmatrix} (-1)^n \exp\!\left[\!\frac{N_0 Q^{-2}\left(\frac{1-\sqrt{1-\epsilon_3}}{2v}\right)}{P_{T} r_1^{-\beta} M_1 \varsigma}   -a n \mu r_1^{-2} \!\right]\! \exp \bigg[-\pi\lambda(r_\mathcal{L}^2 - r_1^2) \sum_{g=1}^G \kappa_g \varrho_g  \bigg( 1- \exp( -a n \mu \varrho_g^{-2}) \nonumber\\
				& \sum_{t=1}^2 \!\!\frac{c_t}{1\!+\!\frac{ Q^{-2}\!\left(\!\frac{1-\sqrt{1-\epsilon_3}}{2v}\!\right)\!}{ r_1^{-\beta} M_1 \varsigma}\!\varrho_g^{-\beta}\! M_t} \!\! \bigg)\! \bigg]\! \exp\!\bigg\{\! \pi \lambda r_\mathcal{L}^2 \!\bigg( \! 1\!- \!\!\sum_{t=1}^2 \frac{c_t}{1\!+\!\frac{ Q^{-2}\!\left(\!\frac{1-\sqrt{1-\epsilon_2}}{2v}\!\right)\! M_t}{ r_1^{-\beta} M_1 \varsigma}} \!\bigg)\!\! - \! \pi \lambda r_\mathcal{L}^2 \!\bigg[ \!\frac{2r_\mathcal{L}^{-\beta }  {}_2F_1\big(\!1,1\!-\!\frac{2}{\beta};2-\frac{2}{\beta}; -\frac{ Q^{-2}\left(\frac{1-\sqrt{1-\epsilon_3}}{2v}\right) M_t r_\mathcal{L}^{-\beta}}{ r_1^{-\beta} M_1 \varsigma}\big) }{\beta-2} \nonumber\\
				&+ \sum_{t=1}^2 \frac{c_t M_t}{\frac{ Q^{-2}\left(\frac{1-\sqrt{1-\epsilon_3}}{2v}\right) M_t}{ r_1^{-\beta} M_1 \varsigma}} \bigg] \bigg\} \bigg\} f_{R_1,R_\mathcal{L}}(r_1,r_\mathcal{L})  \mathrm{d} r_1,r_\mathcal{L}.
			\end{align}
		\end{small}
		\hrulefill
	\end{figure*}
	
	\begin{figure}[t]
		\centerline{\includegraphics[width=0.45\textwidth]{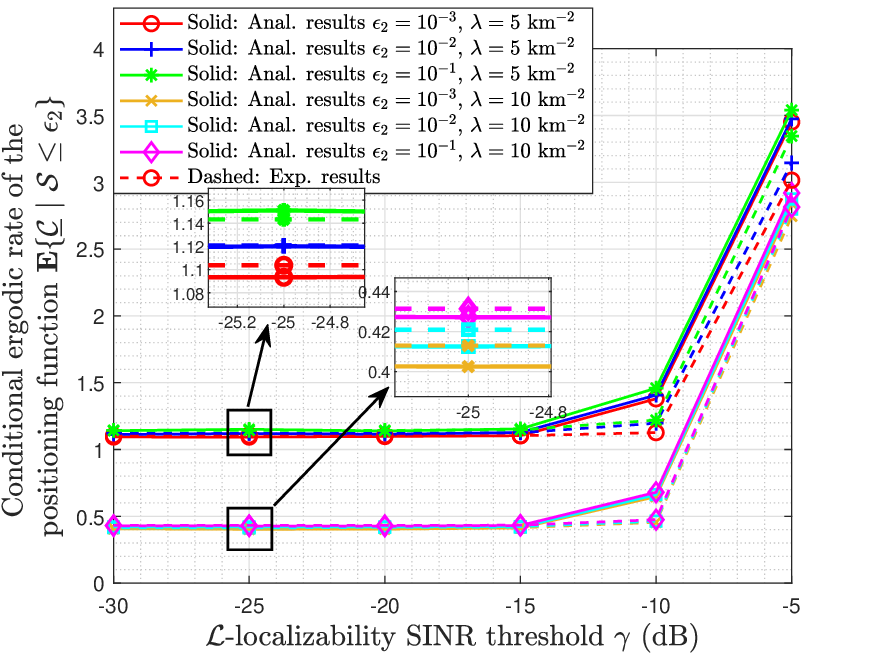}} 
		\caption{The conditional ergodic rate of the positioning function constrained by the communication rate, versus different thresholds of constrained communication rates and BS densities.}\label{fig4_theorem5}
	\end{figure}
	
	\begin{figure}[t]
		\centerline{\includegraphics[width=0.45\textwidth]{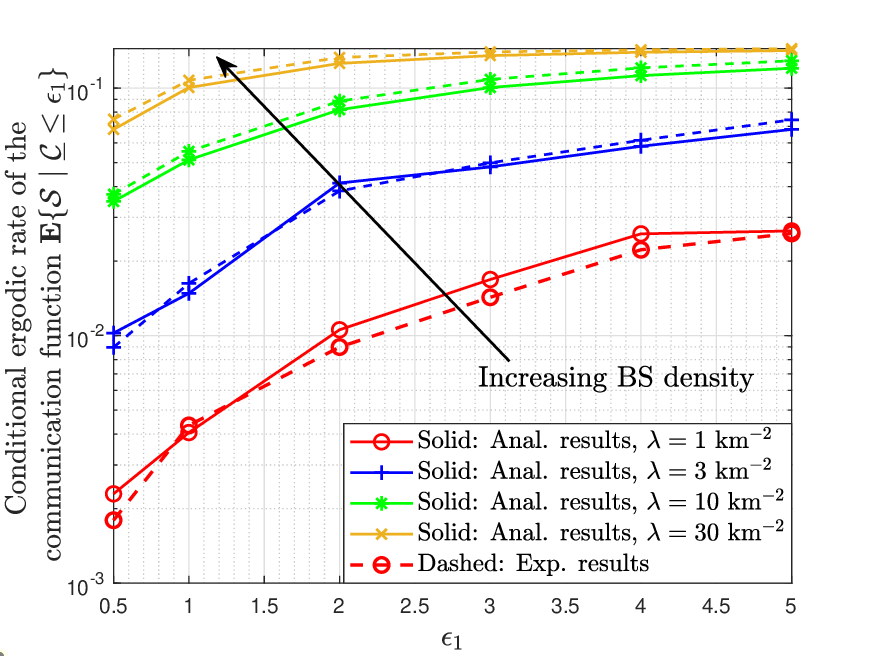}} 
		\caption{The conditional ergodic rate of the communication function constrained by the positioning rate, versus different BS densities.}\label{fig5_theorem5}
	\end{figure}
	
	This result offers useful insights to clarify the design guidelines of network parameters to achieve a satisfactory ISAC coverage rate. In light of this result, we can get the following two results to derive the conditional ISAC coverage rate, indicating the coverage rate of one function in ISAC networks when the other function has bounded limits. Under the condition that the communication SER does not exceed $\epsilon_3$, the coverage rate of the communication process is
	\begin{equation}
		P_{p | s}(\epsilon_1 \ | \ \mathcal{S} \le \epsilon_3) = \frac{P_{p \& s}(\epsilon_1, \epsilon_3)}{P_s(\epsilon_3)},
	\end{equation}
	
	The theoretical result is a very fundamental theoretical basis for the promotion of ISAC networks, which can directly characterize the coverage rate of the positioning function implemented in some existing communication networks. Then, under the condition that the positioning CRLB does not exceed $\epsilon_1$, the coverage rate of the communication process is
	\begin{equation}
		P_{s | p}(\epsilon_3 \ | \ \underline{\mathcal{C}} \le \epsilon_1) = \frac{P_{p \& s}(\epsilon_1, \epsilon_3)}{P_p(\epsilon_1)}.
	\end{equation}
	
	In the following, we provide numerical results to verify and yield useful insights based on the derived theoretical results.

	Fig.~\ref{fig1_3D_2} depicts the coverage rate of ISAC performance for $\lambda=1$ $\text{km}^{-2}$ and $10$ $\text{km}^{-2}$ cases. It can provide many useful insights into the network design to ensure the ISAC dual-function reliability. For instance, the joint probability of sub-$1$ m positioning and sub-$10^{-3}$ SER has increased from $1.4\%$ to $39.8\%$ as the BS density grows from $1$ $\text{km}^{-2}$ to $10$ $\text{km}^{-2}$, indicating a significant enhancement of ISAC coverage in denser networks.

	Fig.~\ref{fig2_theorem5} depicts the trend of the conditional coverage rate of the positioning function under different $\lambda$, given that the communication SER does not exceed $10^{-3}$. It can be noticed that the positioning coverage rate increases significantly with the growth of $\lambda$. Specifically, the conditional coverage rates of sub-$1$ m and sub-$\sqrt{0.5}$ m reach $98.7\%$ and $93\%$, respectively at $\lambda = 50$ km${}^{-2}$, and $68\%$ and $41\%$, respectively at $\lambda=10$ km${}^{-2}$ and $\beta=4.6$.

	Fig.~\ref{fig3_theorem5} presents the conditional coverage rate of the communication function under different $\lambda$ values, given that the positioning CRLB does not exceed $1$. It can be observed that as $\lambda$ increases, the communication performance deteriorates instead, which is an unexpected phenomenon in conventional communication networks. This is mainly due to the fact that, given certain positioning rate regimes, a higher density of BSs leads to shorter distances from the $\mathcal{L}$-nearest BSs to the typical user. Consequently, this results in a notable increase in communication interference, thereby hindering the communication coverage. Besides, it is found that a larger path-loss exponent impaires the conditional coverage rate of the communication function at low $\lambda$ regimes, whereas benefits at high $\lambda$ regimes.

	Fig.~\ref{fig4_theorem5} delves into the conditional ergodic CRLB values under different $\mathcal{L}$-localizability SINR thresholds. It can be found that the conditional ergodic rate of the communication function only has slight improvement when the constrained positioning rate increases. On the other hand, the effect of $\lambda$ is more noticeable. In particular, when $\lambda$ grows from $5$ $\text{km}^{-2}$ to $10$ $\text{km}^{-2}$, the fundamental limit of the conditional rate of the positioning function decreases from $1.07$ m to $0.65$ m. Meanwhile, it is observed that $\mathcal{L}$-localizability SINR threshold has no effect on the conditional ergodic CRLB when it is not larger than $-15$ dB, but deteriorates the positioning performance when $\gamma$ exceeds $-10$ dB.

	Fig.~\ref{fig5_theorem5} provides a straightforward illustration of the conditional ergodic SER performance under different thresholds of the constrained positioning rate. It is shown that the ergodic SER increases from $1.8\times10^{-3}$ to $3.7\times10^{-2}$ when $\lambda$ increases from $1$ $\text{km}^{-2}$ to $10$ $\text{km}^{-2}$ when $\epsilon_1=0.5$, which demonstrates that higher BS density will reduce the conditional communication performance under the constraints of the positioning rates. Meanwhile, the conditional ergodic rate of the positioning function has significant improvement when the constrained communication rate increases.

	\section{CONCLUSIONS}\label{conclusion}
	In this paper, we proposed a generalized stochastic geometry framework to analyze the coverage and ergodic rate performance of ISAC networks. Based on this framework, the coverage rates of sensing and communication performance under resource constraints were defined and calculated. Then, theoretical results for the coverage rate of unified ISAC performance were presented with considering the coupling effects of dual functions in coexistence networks. Further, the analytical formulations for evaluating the ergodic sensing rate constrained by the maximum communication rate, and the ergodic communication rate constrained by the maximum sensing rate were obtained. Numerical results verified the sufficient accuracy of the derived formulations and attained several insights into ISAC network deployment strategies. Specifically, increasing the BS density from $1$ $\text{km}^{-2}$ to $10$ $\text{km}^{-2}$ boosted the ISAC coverage probability from $1.4\%$ to $39.8\%$. Further, results also revealed that with the increase of the constrained maximum sensing rate, the communication rate improves significantly, but the reverse is not obvious.

	\appendix
	\subsection{Preliminaries results 1}\label{eq1}
	The inequality handling $\mathbb{P}(1<c)$, where $c$ is a positive constant, can be addressed by introducing an auxiliary variable $g$ to satisfy the normalized gamma distribution with large parameter $N$ \cite{g_gamma}. Then, the alternative probability $\mathbb{P}(g<c)$ can be tightly upper bounded by
	\begin{equation}
		\mathbb{P}(g<c) \le \left[ 1 - e^{-a c} \right]^N,
	\end{equation}
	where $a = N(N!)^{-\frac{1}{N}}$. It was shown in \cite{mmwave} that this approximation is sufficiently accurate on $N \ge 5$.
	
	\subsection{Preliminaries results 2}\label{eq2}
	For integrals of the form $\int_{\tau_1}^{\tau2} e^{-\mu r^{-2}} r \mathrm{d} r$, we make the following derivation
	\begin{equation}
		\begin{aligned}
			&	\int_{\tau_1}^{\tau2} e^{-\mu r^{-2}} r \mathrm{d} r \overset{\text{a}}{=} \frac{1}{2} \int_{\mu \tau_2^{-2}}^{\mu \tau_1^{-2}} e^{-t} \mathrm{d} \frac{\mu}{t}\\
			& \overset{\text{b}}{=} \frac{\mu}{2} \frac{e^{-t}}{t}|_{\mu \tau_2^{-2}}^{\mu \tau_1^{-2}} + \frac{\mu}{2} \int_{\mu \tau_2^{-2}}^{\mu \tau_1^{-2}} \frac{e^{-t}}{t} \mathrm{d} t \\
			&\overset{\text{c}}{\approx} \frac{\mu}{2} (\frac{e^{-\mu \tau_1^{-2}}}{\mu \tau_1^{-2}} - \frac{e^{-\mu \tau_2^{-2}}}{\mu \tau_2^{-2}} ) + \frac{\mu}{2} \log \frac{1-e^{-b\mu \tau_1^{-2}}}{1-e^{-b\mu \tau_2^{-2}}}  ,   
		\end{aligned}
	\end{equation}
	where (a) is given by letting $t=\mu r^{-2}$; (b) is from computing integration by part; (c) follows from a close approximation with $\int_x^\infty \frac{e^{-t}}{t} \approx - \ln (1-e^{-b x})$ \cite{appro1}.

	\subsection{Proof of Proposition 1}\label{pro1}
	In order to explore the lower bound of the CRLB, we focus on the corresponding the FIM as 
	\begin{equation}\label{J1}
		{J}_{\mathbf{p}_U}= \sum_{l=1}^\mathcal{L} \left( \frac{10 \beta}{\ln 10} \right)^2 r_l^{-2} \xi^{-2}  \begin{bmatrix}
			\cos^2 \theta_l & \cos \theta_l \sin \theta_l \\
			\cos \theta_l \sin \theta_l   &\sin^2 \theta_l
		\end{bmatrix}  .
	\end{equation}
	
	Then the corresponding CRLB can be written as $ \frac{\text{Tr}({J}_{\mathbf{p}_U})}{\text{Det}({J}_{\mathbf{p}_U})}  $.  To minimize the CRLB, it is possible to set the subdiagonal element to zero, which maximizes $\text{Det}({J}_{\mathbf{p}_U})$ while keeping $\text{Tr}({J}_{\mathbf{p}_U})$ unchanged, which leads to
	\begin{equation}\label{con1}
		\sum_{l=1}^L \left( \frac{10 \beta}{\ln 10} \right)^2 r_l^{-2} \xi^{-2} \cos \theta_l \sin \theta_l= 0.
	\end{equation}
	Denote the main diagonal elements in Eq.~\eqref{J1} as $d_1$ and $d_2$. Here the lower bound of the CRLB can be expressed as $\text{CRLB} \ge \frac{d_1+d_2}{d_1 d_2} \ge \frac{4}{d_1+d_2} $, where the second equality is obtained under the condition $d_1=d_2$, i.e.,
	\begin{equation}\label{con2}
		\sum_{l=1}^L \left( \frac{10 \beta}{\ln 10} \right)^2 r_l^{-2} \xi^{-2} \cos^2 \theta_l = \sum_{l=1}^L \left( \frac{10 \beta}{\ln 10} \right)^2 r_l^{-2} \xi^{-2} \sin^2 \theta_l .
	\end{equation} 
	Combining the two lower bound conditions Eq.~\eqref{con1} and Eq.~\eqref{con2}, which are equivalent to
	\begin{equation}\label{con3}
		\sum_{l=1}^L \left( \frac{10 \beta}{\ln 10} \right)^2 r_l^{-2} \xi^{-2} e^{j 2 \theta_l} = 0.
	\end{equation}
	This can be interpreted as the $L$ vectors of length $\left( \frac{10 \beta}{\ln 10} \right)^2 r_l^{-2} \xi^{-2}$ and angle $2 \theta_l$ summing to zero, i.e., these $L$ vectors can be connected head to tail to form an $L$-side shape. This is satisfied by the condition that any sum of $L-1$ edges is greater than the remaining edge. 
	
	\subsection{Proof of Definition 6}\label{de6}
	The proof details of deriving the average achievable rate can refer to \cite{tractable}. Besides, for any positive variable $x$ with conditional PDF $f_X(x| Y \le y)$ and conditional CDF $F_X(x| Y \le y)$, we have the expectation formula as $\mathbb{E}(x| Y \le y)= \int_0^\infty x f_X(x| Y \le y) \mathrm{d} x   = \int_0^\infty \int_0^x \mathrm{d} t f_X(x| Y \le y) \mathrm{d} x=  \int_0^\infty \int_t^\infty f_X(x| Y \le y) \mathrm{d} x \mathrm{d} t = \int_0^\infty (1-F_X(t| Y \le y)) \mathrm{d} t $.
	
	\subsection{Proof of Lemma 2}\label{lemma2}
	The coverge rate of $\mathcal{L}$-conditional positioning performance can be written as
	\begin{equation}
		P_p(\epsilon_1 \ | \ \mathcal{L}) =    \mathbb{P} \left\{  1 \le \mu \sum_{l=1}^{\mathcal{L}} r_l^{-2}  \right\},
	\end{equation}
	where $\mu = \left( \frac{10\beta}{\ln 10} \right)^2 \frac{\epsilon_1}{4\xi^2}$
	This inequality format can be further approximated via Appendix-\ref{eq1} as
	\begin{equation}
		1 + \sum_{n=1}^N \begin{pmatrix}
			N \\ n
		\end{pmatrix} (-1)^n \mathbb{E}_\Phi \left[  e^{-a n \mu \sum_{l=1}^{\mathcal{L}} r_l^{-2}} \right].
	\end{equation}
	
	The expectation part follows from the probability generating functional (PGFL) of the PPP, which states for transformation formula that $\mathbb{E} \left[ \prod_{x \in \Phi} f(x) \right] = \exp\left( -2\pi \lambda \int_r f(r) \right) r \mathrm{d} r$. Although this formula is the Laplace functional of infinite points in a given region, it demonstrates to be quite accurate for the case of finite points in an uncertain region size, following averaging over the region. Thus, it is approximated as
	\begin{equation}
		\begin{aligned}
			&\mathbb{E}_{R_\mathcal{L}} \left\{ \exp\left[ -2\pi \lambda \int_0^{R_\mathcal{L}} \left( 1- e^{-a n \mu r^{-2}} \right) r \mathrm{d} r \right] \right\} = \mathbb{E}_{R_\mathcal{L}} \bigg\{ \\
			&\exp\!\left[\! -\pi \lambda R_\mathcal{L}^2 \! + \! \pi \lambda R_\mathcal{L}^2 e^{-a n \mu R_\mathcal{L}^{-2}} \!\!-\! \pi \lambda \mu \ln\left( 1\!-\! e^{-a n \mu R_\mathcal{L}^{-2}} \right)  \right] \bigg\},
		\end{aligned}
	\end{equation}
	where the equation follows from the derivation of Appendix-\ref{eq2}.
	
	\subsection{Proof of Lemma 3}\label{lemma3}
	The $\mathcal{L}$-localizability can be transformed to $\mathbb{P}\left\{ 1 \ge \gamma r_\mathcal{L}^\beta \sum_{i=\mathcal{L}+1}^\infty r_i^{-\beta} + \gamma N_0 P_T^{-1} r_\mathcal{L}^\beta \right\}$.
	
	Based on Appendix-\ref{eq1}, it can be rewritten as
	\begin{equation}
		\sum_{n=1}^N \begin{pmatrix}
			N \\ n
		\end{pmatrix} (-1)^{n+1} e^{a n \gamma r_\mathcal{L}^\beta N_0 P_T^{-1}} \mathbb{E}_\Phi \left[  e^{-a n \gamma r_\mathcal{L}^\beta \sum_{i=\mathcal{L}+1}^\infty r_i^{-\beta}} \right],
	\end{equation}
	where the expectation part can be obtained by
	\begin{equation}
		\begin{aligned}
			&\exp\left[ -2\pi\lambda \int_{r_\mathcal{L}}^\infty \left( 1-e^{-a n \gamma r_\mathcal{L}^\beta r^{-\beta}} \right) r \mathrm{d} r \right] \\
			&=\exp\left\{ \pi \lambda r_\mathcal{L}^2 \left[  1-e^{-a n \gamma }+ (a n \gamma)^{\frac{2}{\beta}} \Gamma(-\frac{2}{\beta}, 0, a n \gamma) \right] \right\},
		\end{aligned}
	\end{equation}
	wherein $\Gamma(a,z_0,z_1)$ is the generalized incomplete gamma function as $\int_{z_0}^{z_1} t^{a-1} e^{-t} \mathrm{d} t$.
	
	\subsection{Proof of Theorem 2}\label{theo2}
	{With the positioning process, the BS will align the beam towards the nearest user to exploit the maximum directivity gain, i.e., $M_1$. Then, we proceed with the coverage of the communication process by deriving	  	\vspace{-7pt}
	\begin{equation}
		\begin{aligned}
			& \mathbb{P}\left\{ \frac{P_{T} r_1^{-\beta} |\alpha_1|^2 M_1 }{\mathcal{I}_{agg}+\sigma_n^2} \ge \epsilon_2   \right\}  =  \mathbb{P}\left\{ |\alpha_1|^2 \ge \frac{\epsilon_2 (\mathcal{I}_{agg}+\sigma_n^2)}{P_{T} r_1^{-\beta} M_1 }  \right\} \\
			&  \overset{\text{(a)}}{=}\mathbb{E}_\Phi \!\! \left\{ \!\! \frac{\epsilon_2 (\mathcal{I}_{agg}+\sigma_n^2)}{P_{T} r_1^{-\beta} M_1 } \!\!\right\} \!\!=e^{-\frac{\epsilon_2 \sigma_n^2}{P_{T} r_1^{-\beta} M_1 }  }  \mathcal{L}_{\mathcal{I}_{agg}}\!\!\left(\!\! \frac{ \epsilon_2}{P_{T} r_1^{-2\beta} M_1 } \!\!\right)\!\! ,
		\end{aligned}
	\end{equation}
	where $\mathcal{I}_{agg} = \sum_{k =2}^\infty P_{T} r_k^{-\beta} |\alpha_k|^2 D_k $ and (a) is the CDF of $ |\alpha_1|^2$ as an exponential distribution with parameter 1. As such, considering the average small-scale fading $|\alpha|^2$ and directional beam gain $D$ in the PGFL of the aggregate interference power, we have $\mathcal{L}_{\mathcal{I}_{agg}}(s)$	  	\vspace{-7pt}
	\begin{equation}
		\begin{aligned}
			&= \exp \left\{ - 2\pi \lambda \int_{r_1}^{\infty}\!\! \left[ 1-  \mathbb{E}_{\alpha, D} \{ \exp(-s P_T r^{-\beta}|\alpha|^2 D) \} \right] r \mathrm{d} r   \right\} \\
			& = \exp \left\{ - 2\pi \lambda \int_{r_1}^{\infty} \left[ 1-  \sum_{t=1}^2 \frac{c_t}{1+sP_Tr^{-\beta} M_t}   \right] r \mathrm{d} r   \right\} .
		\end{aligned}
	\end{equation}}
	
	{Using the method of integration by parts, we have $\frac{1}{2} \left(  1-  \sum_{t=1}^2 \frac{c_t}{1+sP_Tr^{-\beta} M_t}   \right) r^2 \big|_{r_1}^{\infty} + \frac{\beta}{2} \int_{r_1}^\infty r^2 \sum_{t=1}^2  \frac{s P_T c_t M_t r^{-\beta-1}}{(1+sP_T r^{-\beta} M_t)^2}$. Define the first part as $f(r) = \frac{1}{2} \left(  1-  \sum_{t=1}^2 \frac{c_t}{1+sP_Tr^{-\beta} M_t}   \right) r^2$, we have $h(r) \le f(r) \le g(r)$, where $h(r) = \frac{1}{2} \left(  1-  \frac{1}{1+sP_Tr^{-\beta} M_2}   \right) r^2$ and $g(r) =  \frac{1}{2} \left(  1-  \frac{1}{1+sP_Tr^{-\beta} M_1}   \right) r^2$. It is readily to prove that when $\beta>1$, there exists $\lim_{r \to \infty} h(r) = \lim_{r \to \infty} g(r) = 0$.}
	
	{By means of the Squeeze Theorem, we get the result $\lim_{r \to \infty} f(r) = 0$. Then, the first part can be further written as $-\frac{1}{2} \left(  1-  \sum_{t=1}^2 \frac{c_t}{1+sP_T M_t}   \right) r_1^{\beta+2}$. On the other hand, the second part can be obtained as 
	\begin{equation}
		\begin{aligned}
			&\frac{s P_T r_1^2}{2}\bigg[  \sum_{t=1}^2 \frac{c_t M_t}{s P_t M_t + r_1^\beta  } + \frac{2 r_1^{-\beta}}{\beta-2} \ {}_2F_1(1,1-\frac{2}{\beta}; \\
			&2-\frac{2}{\beta};-sP_T M_T r_1^{-\beta}) \bigg].
		\end{aligned}
	\end{equation}
}
	\subsection{Proof of Theorem 3}\label{theo3}
	{The coverage rate of ISAC performance in \emph{Definition 5} can be written as Eq.~\eqref{theo3_1}, which breaks down into two parts, i.e., $\Lambda_1$ and $\Lambda_2$, wherein $\Lambda_1$ has been derived in Appendix-\ref{theo2} as}
	\begin{figure*}[t]
	{	\begin{align}\label{theo3_1}
			&\mathbb{P}\left\{  |\alpha_1|^2 \ge \frac{\epsilon_2 (\mathcal{I}_{agg}+\sigma_n^2)}{P_{T} r_1^{-\beta} M_1 } , \ g < \left( \frac{10\beta}{\ln 10} \right)^2 \frac{\epsilon_1}{4\xi^2} \sum_{l=1}^{\mathcal{L}} r_l^{-2} \ \bigg| \ \mathcal{L} \right\} \notag \\
			& =\mathbb{E}_\Phi \left\{ e^{\frac{\epsilon_2 (\mathcal{I}_{agg}+\sigma_n^2)}{P_{T} r_1^{-\beta} M_1 }} \left[ 1 + \sum_{n=1}^N \begin{pmatrix}
				N \\ n
			\end{pmatrix} (-1)^n  e^{-a n \mu \sum_{l=1}^{\mathcal{L}} r_l^{-2}} \right] \ \bigg| \ \mathcal{L}  \right\} \notag \\
			& = \begin{matrix} \underbrace{ \mathbb{E}_\Phi\left\{e^{\frac{\epsilon_2 (\mathcal{I}_{agg}+\sigma_n^2)}{P_{T} r_1^{-\beta} M_1 }}  \right\}  } \\ \Lambda_1  \end{matrix} +  \sum_{n=1}^N \begin{pmatrix}
				N \\ n
			\end{pmatrix} (-1)^n  \begin{matrix} \underbrace{ \mathbb{E}_\Phi \left\{ e^{\frac{\epsilon_2 (\mathcal{I}_{agg}+\sigma_n^2)}{P_{T} r_1^{-\beta} M_1 }-a n \mu \sum_{l=1}^{\mathcal{L}} r_l^{-2}}  \ \bigg| \ \mathcal{L} \right\}} \\ \Lambda_2 \end{matrix} 
		\end{align}}
		\hrulefill
	\end{figure*}
	
	\begin{equation}
		\Lambda_1 = e^{-\frac{\epsilon_2 \sigma_n^2}{P_{T} r_1^{-\beta} M_1 }  }  \mathcal{L}_{\mathcal{I}_{agg}}\left( \frac{ \epsilon_2}{P_{T} r_1^{-2\beta} M_1 } \right) .
	\end{equation}
	
	{Further, we split the $\Lambda_2$ into three conditionally independent parts under the conditions $(R_1, \ R_\mathcal{L})$ as
	\begin{small}
		\begin{equation}
			\begin{aligned}
				&\mathbb{E}_\Phi \bigg\{ \begin{matrix} \underbrace{\exp\left\{\frac{\epsilon_2 \sigma_n^2 }{P_{T} r_1^{-\beta} M_1 \varsigma}   -a n \mu r_1^{-2} \right\} } \\ \Lambda_{2,1} \end{matrix} \\
				&  \begin{matrix} \underbrace{\cdot \exp\left\{ \sum_{l=2}^\mathcal{L} \frac{\epsilon_2 P_{T}  |\alpha_l|^2 D_l }{P_{T} r_1^{-\beta} M_1}r_l^{-\beta}   -a n \mu r_l^{-2} \right\}  } \\ \Lambda_{2,2} \end{matrix} \\
				& \begin{matrix} \underbrace{ \cdot \exp\left\{\sum_{k=\mathcal{L}+1}^\infty \frac{\epsilon_2 P_{T}  |\alpha_k|^2 D_k }{P_{T} r_1^{-\beta} M_1 }r_k^{-\beta} \right\}  } \\ \Lambda_{2,3} \end{matrix} \ \bigg| \ (R_1, \ R_\mathcal{L}, \ \mathcal{L}) \bigg\} .
			\end{aligned}
		\end{equation}
	\end{small}}
	
	{We can approximate the second part $\Lambda_{2,2}$ by transforming it into the expectation $\{R_1, R_\mathcal{L}, \mathcal{L}\}$ on
	\begin{small}
		\begin{equation}
			\begin{aligned}
				&\exp \!\!\bigg[ \!\!-2\pi\lambda \!\int_{r_1}^{r_\mathcal{L}} \!\! \bigg(\! 1- \mathbb{E}_{|\alpha|^2, D}\bigg( \!\!\exp\!\! \bigg(\frac{\epsilon_2  |\alpha|^2 D }{ r_1^{-\beta} M_1} r^{-\beta}   -a n \mu r^{-2}\bigg) \! \bigg) \!\bigg)\\
				& r \mathrm{d} r \bigg] \!= \!\exp\!\! \bigg[\!\!-2\pi\lambda\!\! \int_{r_1}^{r_\mathcal{L}} \!\!\bigg( \!\!1\!-\! \sum_{t=1}^2 \frac{c_t}{1+\frac{\epsilon_2 r^{-\beta} M_t }{ r_1^{-\beta} M_1 }} \exp( -a n \mu r^{-2})\! \bigg) r \mathrm{d} r.
			\end{aligned}
		\end{equation}
	\end{small}}
	
	{This expression can be simplified using Gaussian Quadrature rules \cite{quadrature} as $\exp \bigg[-\pi\lambda(r_\mathcal{L}^2 - r_1^2) \sum_{g=1}^G \kappa_g \varrho_g  \bigg( 1- \sum_{t=1}^2 \frac{c_t}{1+\frac{ \epsilon_2  \varrho_g^{-\beta} M_t}{ r_1^{-\beta} M_1 }} \exp( -a n \mu \varrho_g^{-2}) \bigg)  \bigg]$, where $\varrho_g$ and $\kappa_g$ are the $g$-th variable and weighting factor of the Laguerre polynomials, respectively.}
	
	{Finally, the third part $\Lambda_{2,3}$ can be transformed by PGFL to
	\begin{small}
		\begin{equation}
			\begin{aligned}
				&\exp\bigg\{ - 2\pi \lambda \int_{r_\mathcal{L}}^\infty  \bigg(  1-  \mathbb{E}_{|\alpha|^2, D}\bigg( \exp\bigg(\frac{\epsilon_2  |\alpha|^2 D }{ r_1^{-\beta} M_1 } \bigg) r^{-\beta} \bigg) \bigg\}\\
				& = \exp\bigg\{ - 2\pi \lambda \int_{r_\mathcal{L}}^\infty  \bigg(  1- \sum_{t=1}^2 \frac{c_t}{1+\frac{ \epsilon_2  r^{-\beta} M_t}{ r_1^{-\beta} M_1 }}   \bigg)  r \mathrm{d} r \bigg\} \\
				&= \exp\bigg\{ \pi \lambda r_L^2\bigg(  1- \sum_{t=1}^2 \frac{c_t}{1+\frac{ \epsilon_2 M_t}{ r_1^{-\beta} M_1 \varsigma}} \bigg) - \pi \lambda r_\mathcal{L}^2 \\
				&\bigg[ \frac{2r_\mathcal{L}^{-\beta }  {}_2F_1\big(1,1-\frac{2}{\beta};2-\frac{2}{\beta}; -\frac{ \epsilon_2 M_t r_\mathcal{L}^{-\beta}}{ r_1^{-\beta} M_1 }\big) }{\beta-2} + \sum_{t=1}^2 \frac{c_t  r_1^{-\beta} M_1 }{\epsilon_2} \bigg] \bigg\}.
			\end{aligned}
		\end{equation}
	\end{small}}
	
	{The expectation for $(R_1, \ R_\mathcal{L}, \ \mathcal{L})$ is to multiply by the corresponding PDF or PMF.}


\begin{thebibliography}{11}
		\bibitem{ISAC1}
		F. Dong, F. Liu, Y. Cui, W. Wang, K. Han and Z. Wang, ``Sensing as a service in 6G perceptive networks: A unified framework for ISAC resource allocation,'' \emph{IEEE Trans. Wireless Commun.}, vol. 22, no. 5, pp. 3522-3536, May 2023.
		
		\bibitem{ISAC1_1}
		Z. Wei \emph{et al.}, ``Integrated sensing and communication signals toward 5G-A and 6G: A survey,'' \emph{IEEE IEEE Internet Things J.}, vol. 10, no. 13, pp. 11068-11092, 1 Jul. 2023.
		
		\bibitem{application}
		J. A. Zhang \emph{et al.}, ``Enabling joint communication and radio sensing in mobile networks—A survey,'' \emph{IEEE Commun. Surveys Tut.}, vol. 24, no. 1, pp. 306–345, 2021.
		
		\bibitem{ISAC2}
		F. Liu \emph{et al.}, ``Integrated sensing and communications: Toward dual-functional wireless networks for 6G and beyond,'' \emph{IEEE J. Sel. Areas Commun.}, vol. 40, no. 6, pp. 1728-1767, Jun. 2022.
		
		\bibitem{ISAC3}
		C. Zhang, W. Yi, Y. Liu and L. Hanzo, ``Semi-integrated-sensing-and-communication (Semi-ISaC): From OMA to NOMA," \emph{IEEE Trans. Commun.}, vol. 71, no. 4, pp. 1878-1893, Apr. 2023.
		
		\bibitem{ISAC4}
		W. Zhou, R. Zhang, G. Chen and W. Wu, ``Integrated sensing and communication waveform design: A survey,'' \emph{IEEE Open J. Commun. Soc.}, vol. 3, pp. 1930-1949, Oct. 2022.
		
		\bibitem{ISAC5}
		X. Gan, C. Huang, Z. Yang, C. Zhong, X. Chen, Z. Zhang, Q. Guo, C. Yuen, M. Debbah, ``Bayesian learning for double-RIS aided ISAC systems with superimposed pilots and data'', 2024, \emph{arXiv: 2402.10593}, [Online]. Available: https://arxiv.org/abs/2402.10593
		
		\bibitem{ISAC6}
		X. Tong, Z. Zhang, J. Wang, C. Huang and M. Debbah, ``Joint multi-user communication and sensing exploiting both signal and environment sparsity,'' \emph{IEEE J. Sel. Topics Signal Process.}, vol. 15, no. 6, pp. 1409-1422, Nov. 2021.
		
		\bibitem{chen}
		M. Chen, D. Gündüz, K. Huang, W. Saad, M. Bennis, A. V. Feljan, and H. V. Poor, ``Distributed learning in wireless networks: Recent progress and future challenges," \emph{IEEE J. Sel. Areas Commun.}, vol. 39, no. 12, pp. 3579-3605, Dec. 2021.
		
		\bibitem{wang}
		D. Wang et al., ``Mean field game-based waveform precoding design for mobile crowd integrated sensing, communication, and computation systems,'' \emph{IEEE Trans. Wireless Commun.}, early access, Mar. 2024, doi: 10.1109/TWC.2024.3372033.
		
		
		\bibitem{ISAC_performance}
		A. Liu \emph{et al.}, ``A survey on fundamental limits of integrated sensing and communication,'' \emph{IEEE Commun. Surveys Tuts.}, vol. 24, no. 2, pp. 994-1034, 2nd quarter 2022.
		
		\bibitem{theo_ISAC1}
		Y. Xiong, F. Liu, Y. Cui, W. Yuan, T. X. Han and G. Caire, ``On the fundamental tradeoff of integrated sensing and communications under Gaussian channels,'' \emph{IEEE Trans. Inform. Theory}, vol. 69, no. 9, pp. 5723-5751, Sept. 2023.
		
		\bibitem{theo_ISAC2}
		C. Ouyang, Y. Liu, H. Yang and N. Al-Dhahir, ``Integrated sensing and communications: A mutual information-based framework,'' \emph{IEEE Commun. Mag.}, vol. 61, no. 5, pp. 26-32, May 2023.
		
		\bibitem{nearfield}
		X. Gan, C. Huang, Z. Yang, C. Zhong and Z. Zhang, ``Near-field localization for holographic RIS assisted mmWave systems,'' \emph{IEEE Commun. Lett.}, vol. 27, no. 1, pp. 140-144, Jan. 2023.
		
		\bibitem{mmwave}
		T. Bai and R. W. Heath, ``Coverage and rate analysis for millimeter-wave cellular networks," \emph{IEEE Trans. Wireless Commun.}, vol. 14, no. 2, pp. 1100-1114, Feb. 2015.
		
		\bibitem{SG2}
		H. ElSawy, \emph{et al.}, ``Modeling and analysis of cellular networks using stochastic geometry: A tutorial,'' \emph{IEEE Commun. Surveys Tuts}, vol. 19, no. 1, pp. 167-203, 2nd Quart., 2017.
		
		
		\bibitem{Gil}
		M. Di Renzo and P. Guan, ``Stochastic geometry modeling of coverage and rate of cellular networks using the Gil-Pelaez inversion theorem,'' \emph{IEEE Commun. Lett.}, vol. 18, no. 9, pp. 1575-1578, Sept. 2014.
		
		\bibitem{SG1}
		M. Haenggi, \emph{Stochastic Geometry for Wireless Networks}. Cambridge, U.K.: Cambridge Univ. Press, 2012.
		
		\bibitem{SG3}
		Y. Hmamouche, M. Benjillali, S. Saoudi, H. Yanikomeroglu and M. D. Renzo, ``New trends in stochastic geometry for wireless networks: A tutorial and survey,'' \emph{Proc. IEEE}, vol. 109, no. 7, pp. 1200-1252, Jul. 2021.
		
		\bibitem{l-localizability}
		J. Schloemann, H. S. Dhillon, and R. M. Buehrer, ``Toward a tractable analysis of localization fundamentals in cellular networks,'' \emph{IEEE Trans. Wireless Commun.}, vol. 15, no. 3, pp. 1768-1782, Mar. 2016.
		
		\bibitem{hearability}
		J. Schloemann, H. S. Dhillon and R. M. Buehrer, ``A tractable analysis of the improvement in unique localizability through collaboration,'' \emph{IEEE Trans. Wireless Commun.}, vol. 15, no. 6, pp. 3934-3948, Jun. 2016.
		
		\bibitem{SG4}
		C. E. O’Lone, H. S. Dhillon and R. M. Buehrer, ``A statistical characterization of localization performance in wireless networks,'' \emph{IEEE Trans. Wireless Commun.}, vol. 17, no. 9, pp. 5841-5856, Sept. 2018.
		
		\bibitem{SG5}
		J. He, Y. J. Chun and H. C. So, ``A unified analytical framework for RSS-based localization systems,'' \emph{IEEE Internet Things J.}, vol. 9, no. 9, pp. 6506-6519, May 2022.
		
		\bibitem{crlb}
		S. M. Kay, \emph{Fundamentals of Statistical Signal Processing, Volume I: Estimation Theory} (Prentice-Hall Signal Processing Series). Englewood Cliffs, NJ, USA: Prentice-Hall, 1993.
		
		\bibitem{crlb2}
		Z. Yu, X. Hu, C. Liu, M. Peng, C. Zhong, ``IRS-aided non-orthogonal ISAC systems: Performance analysis and beamforming design'', 2022, \emph{arXiv: 2208.05324}, [Online]. Available: https://arxiv.org/abs/2208.05324
		
		\bibitem{crlb3}
		B. Guo, J. Liang, B. Tang, L. Li and H. C. So, ``Bistatic MIMO DFRC system waveform design via symbol distance/direction discrimination,'' \emph{IEEE Trans. on Signal Process.}, vol. 71, pp. 3996-4010, Oct. 2023.
		
		
		
		\bibitem{SG_ISAC1}
		N. R. Olson, J. G. Andrews, Jr R. W. Heath, ``Coverage and capacity of joint communication and sensing in wireless networks.'', 2022, \emph{arXiv: 2210.02289}, [Online]. Available: https://arxiv.org/abs/2210.02289
		
		\bibitem{SG_ISAC2}
		C. Skouroumounis, C. Psomas, and I. Krikidis, ``FD-JCAS techniques for mmWave HetNets: Ginibre point process modeling and analysis,'' \emph{IEEE Trans. Mob. Comput.}, vol. 21, no. 12, pp. 4352-4366, 1 Dec. 2022.
		
		\bibitem{SG_ISAC3}
		S. S. Ram, G. Singh, and G. Ghatak, ``Optimization of radar parameters for maximum detection probability under generalized discrete clutter conditions using stochastic geometry,'' \emph{IEEE Open J. Signal Process.}, vol. 2, pp. 571–585, 2021.
		
		\bibitem{SG_ISAC4}
		P. Ren, A. Munari, and M. Petrova, ``Performance tradeoffs of joint radar-communication networks,'' \emph{IEEE Wireless Commun. Lett.}, vol. 8, no. 1, pp. 165-168, Aug. 2019.
		
		\bibitem{SG_ISAC5}
		K. Meng, C. Masouros, G. Chen, F. Liu, ``Network-level integrated sensing and communication: Interference management and BS coordination using stochastic geometry.'' 2023, \emph{arXiv: 2311.09052}, [Online]. Available: https://arxiv.org/abs/2311.09052
		
		\bibitem{benefit}
		Y. Guo, C. Yin, O. Lu, M. Wang and B. Xia, ``Performance analysis for mm-Wave ISAC systems with mutual benefit,'' in \emph{Proc. IEEE Glob. Commun. Conf. (GLOBECOM)}, Rio de Janeiro, Brazil, Dec. 2022, pp. 5438-5443.
		
		\bibitem{gan}
		X. Gan, C. Huang, Z. Yang, C. Zhong, X. Chen, Z. Zhang, Q. Guo, C. Yuen, M. Debbah, ``Simultaneous communication and localization for double-RIS aided multi-UE ISAC systems,'' in \emph{Proc. IEEE 23rd Int. Conf. Commun. Technol. (ICCT)}, Wuxi, China, 2023, pp. 422-427.
		
		
		\bibitem{BPP}
		M. Afshang and H. S. Dhillon, ``Fundamentals of modeling finite wireless networks using binomial point process'', \emph{IEEE Trans. Wireless Commun.}, vol. 16, no. 5, pp. 3355-3370, May 2017.
		
		
		\bibitem{g_gamma}
		R. Aris, \emph{Mathematical Modeling: A Chemical Engineer’s Perspective.} New York, NY, USA: Academic, 1999.
		
		\bibitem{appro1}
		H. Alzer, ``On some inequalities for the incomplete Gamma function,'' \emph{Math. Comput.}, vol. 66, no. 218, pp. 771-778, Apr. 1997.
		
		\bibitem{tractable}
		J. G. Andrews, F. Baccelli, and R. Krishna Ganti, ``A tractable approach to coverage and rate in cellular networks,'' \emph{IEEE Trans. Commun.}, vol. 59, no. 11, pp. 3122–3134, Nov. 2011.
		
		
		\bibitem{quadrature}
		M. Abramowitz and I. A. Stegun, \emph{Handbook of Mathematical Functions With Formulas, Graphs, and Mathematical Tabl}, Washington,D.C.: U.S. Dept. Commerce, 1972.
		
		
		
	\end{thebibliography}
\end{document}